\documentclass[journal,twoside,web]{ieeecolor}
\usepackage{generic}
\usepackage{cite}
\usepackage{framed,multirow,booktabs}
\usepackage{amsmath,amssymb,amsfonts}
\usepackage{algorithmic}
\usepackage{graphicx}
\usepackage{algorithm,algorithmic}
\usepackage{hyperref}
\hypersetup{hidelinks=true}
\usepackage{textcomp}
\usepackage{bm}
\usepackage{epstopdf}
\usepackage{array}
\usepackage{epsfig}
\usepackage{times}
\usepackage{tabularx}
\usepackage{float}

\def\BibTeX{{\rm B\kern-.05em{\sc i\kern-.025em b}\kern-.08em
    T\kern-.1667em\lower.7ex\hbox{E}\kern-.125emX}}
\begin{document}
\title{A Symmetric Dynamic Learning Framework for Diffeomorphic Medical Image Registration}
\author{Jinqiu Deng, Ke Chen, Mingke Li, Daoping Zhang, Chong Chen, Alejandro F. Frangi \IEEEmembership{Fellow, IEEE}, and Jianping Zhang
\thanks{This work was supported in part by the National Natural Science Foundation of China (12071402, 12201320), the science and technology innovation Program of Hunan Province (2024WZ9008), the Natural Science Foundation of Hunan Province (2023GK2029), the Fundamental Research Funds for Central Universities (63221039, 63231144), the Royal Academy of Engineering Chair INSILEX (CiET1819\textbackslash19), the UKRI Frontier Research Guarantee INSILICO (EP\textbackslash Y030494\textbackslash 1), and the NIHR Manchester Biomedical Research Centre.} 
\thanks{Corresponding authors: J. P. Zhang (jpzhang@xtu.edu.cn), A. F. Frangi (alejandro.frangi@manchester.ac.uk).}
\thanks{J. Q. Deng, M. K. Li and J. P. Zhang are with the School of Mathematics and Computational Science, Xiangtan University, Xiangtan, China.}
\thanks{K. Chen is with the Department of Mathematics and Statistics, University of Strathclyde, Glasgow, UK.}
\thanks{D. P. Zhang is with the Department of Mathematical Sciences, Nankai University, Tianjin, China.}
\thanks{C. Chen is with the Academy of Mathematics and Systems Science, Chinese Academy of Sciences, 
Beijing, China.}
\thanks{A. F. Frangi is with the Schools of Health Sciences and Computer Science, University of Manchester, Manchester, UK. }
}

\maketitle

\begin{abstract}
Diffeomorphic image registration is crucial for various medical imaging applications because it can preserve the topology of the transformation. This study introduces DCCNN-LSTM-Reg, a learning framework that evolves dynamically and learns a symmetrical registration path by satisfying a specified control increment system. This framework aims to obtain symmetric diffeomorphic deformations between moving and fixed images. To achieve this, we combine deep learning networks with diffeomorphic mathematical mechanisms to create a continuous and dynamic registration architecture, which consists of multiple Symmetric Registration (SR) modules cascaded on five different scales. Specifically, our method first uses two U-nets with shared parameters to extract multiscale feature pyramids from the images. We then develop an SR-module comprising a sequential CNN-LSTM architecture to progressively correct the forward and reverse multiscale deformation fields using control increment learning and the homotopy continuation technique. Through extensive experiments on three 3D registration tasks, we demonstrate that our method outperforms existing approaches in both quantitative and qualitative evaluations.

\end{abstract}


\section{Introduction}\label{sec:introduction}
\IEEEPARstart{D}{eformable} image registration is a crucial technique in medical image analysis to align anatomical structures in images \cite{YFu1}. This technique is essential for various clinical applications, including lesion identification \cite{Yang2012}, dose accumulation \cite{Velec2011}, motion tracking \cite{Fu2011}, and image reconstruction \cite{Dang2014}. Traditional registration methods typically formulate image registration as a variational problem and solve it iteratively using optimization algorithms \cite{JPZhang2015}, such as Demons \cite{demons1998}, B-spline \cite{B1999}, LDDMM \cite{LDDMM2005}, Diffeomorphic Demons \cite{diffdemons2009}, SyN \cite{syn2008}, diffeomorphic image registration with control increment constraint \cite{ZHANG2022}, and their variants \cite{KCLam2014, CChen2018,Daoping2018, CChen2019,HHan2020a, Chen_2021,HHan2021a, DZhang2022}. Although these approaches preserve diffeomorphism and offer high registration accuracy, they are computationally expensive and slow because not only the time-dependent sequence operations but tuning hyper-parameter are needed for each image pair.

As AlexNet \cite{Alom2018} achieved success in ImageNet challenge, deep learning algorithms have been increasingly used in various image processing applications, achieving remarkable results in most tasks. In recent years, there have been many deep learning frameworks to solve medical image registration problems \cite{YFu1, Haskins2020,PXue2024Str}. Initially, training neural networks requires the supervision of ground-truth deformation fields. Recently, unsupervised learning techniques employing a convolutional neural network (CNN), particularly U-net, have become the main focus of research in deep learning registration algorithms \cite{VoxelMorph2019,diffVoxelMorph2018,vtn2019, CycleMorph2020,sym2020, RLiu2022,prnet2022, Wei2022,YLiu2023Geo, AHering2023}. Unlike traditional methods, unsupervised deep learning registrations have remarkably improved computational speed while maintaining accuracy \cite{CLiu2024Reg}. 

Current learning methodologies, such as VoxelMorph \cite{VoxelMorph2019}, use two concatenated images as input and apply the U-net architecture to directly extract features, then generate deformation or velocity fields. However, we discuss that these straightforward methods may lack accuracy in complex scenarios. For complex or large-scale deformations, the VTN framework \cite{vtn2019}, which uses a cascade of multiple networks, proves to be an effective approach. Typically, these cascades consist of serially connected U-nets, where each progressively learns the deformation field and transforms the moving image to align with the fixed image through interpolation. However, this method involves high computational costs and tends to overfitting. It also accumulates errors during multiple interpolations, making it challenging to maintain the diffeomorphism. Moreover, most existing deep learning techniques are limited to unidirectional registration, neglecting the invertibility property of the smooth deformation field. Although SYM-net \cite{sym2020} and similar approaches have explored symmetrical registration, they still rely on a single U-net to learn spatial transformations and do not integrate cascaded and symmetrical registration.

The scaling and squaring method \cite{Arsigny2006} is widely adopted for diffeomorphic registration. However, these techniques are limited by the assumption of a constant velocity field, which may constrain their capability to capture fine-scale deformations. Additionally, the coupling nature of their iterative solutions can lead to interpolation errors, challenging practical application. Traditional methods are often considered more efficient than deep learning techniques in preserving diffeomorphism. Certain methods provide theoretical guarantees for diffeomorphic registration without requiring repeated interpolation. Zhang and Li \cite{ZHANG2022} examined the optimal control relaxation method to indirectly determine the diffeomorphic transformation through the Jacobian determinant equation and investigated its applications in medical image registration. They developed the final deformation field by progressively incorporating control increment sequences that satisfy a particular PDE system into the previous deformation field. This inspired us to design a registration network that incorporates multiple incremental fields at different stages to compute the deformation field.

This study investigates the mathematical diffeomorphic mechanisms proposed by Zhang and Li \cite{ZHANG2022} and formulates diffeomorphic registration as a dynamic system. To address this system with a learning-based approach, we explore the long-term memory capabilities of the LSTM network to facilitate integrated multiscale cascade architecture and symmetrical registration path, and then propose a Diffeomorphic Cascaded CNN-LSTM Registration (DCCNN-LSTM-Reg) framework. This framework utilizes two U-nets that share the same parameters to extract multiscale features from a pair of images. We then develop an SR-module comprising a sequential CNN-LSTM architecture to iteratively align the images from coarse to fine levels using control increment learning and the homotopy continuation method. The suggested SR-module integrates a symmetric registration path based on its reversibility to further improve the performance of the progressive registration. In addition, we used intermediate deformation fields to progressively register the extracted features at subsequent finer scales, refining the registration accuracy. The main contributions of this work are summarized as follows:
\begin{itemize}
\item \textbf{Dynamical deformation framework:} We model diffeomorphic image deformation using a dynamical system with control increments. Using homotopy continuation, we integrate all multiscale incremental fields to learn the evolving trajectory of diffeomorphic deformation fields. This technique enables us to obtain more flexible and accurate deformation fields.
\item \textbf{Enhanced cascaded CNN-LSTM architecture:} We propose a modified CNN-LSTM control increment module for cascaded transformation correction to achieve diffeomorphic multiscale registration. The proposed CNN-LSTM structure has advantages in capturing long-term dependencies of cascaded diffeomorphic registration.
\item \textbf{Symmetric diffeomorphic registration:} In the SR-module, we establish two symmetric registration paths with shared parameters which simultaneously generate symmetric deformation fields by reversing the order of input features. SR-module allows the invertibility of registration to be incorporated into the learning framework through optimizing the cyclic consistency loss, and not only yields symmetric deformation fields but also ensures diffeomorphism.
\item \textbf{Pre-align of features:} The deformation field obtained from the previous cascade is used to pre-align features of both the moving and fixed images in the subsequent cascade, thus increasing the accuracy of the registration.
\end{itemize}

\section{Related work}
\label{section:2}
This section provides a brief overview of model-based and data-driven approaches to image registration.
\subsection{Diffeomorphic Image Registration}
The most challenging type of medical image registration is deformable image registration (DIR), especially diffeomorphic DIR. When optimizing an energy function, diffeomorphic DIR establishes the spatial transformation relationship between two images. Let $X$ and $Y$ represent the moving and fixed images, the diffeomorphic deformation field $\hat{\bm{\phi}}$ of image registration is determined by minimizing an energy function as
\begin{equation}\label{eqdiff}
\begin{split}
\hat{\bm{\phi}} = \mathop {\arg\rm{min}}\limits_{\bm{\phi}  \in \text{Diff}(\Omega)} \mathcal{L}_{sim}(X,Y \circ \bm{\phi}) + \lambda \mathcal{L}_{smooth}(\bm{\phi}),
\end{split}
\end{equation}
where $\text{Diff}(\Omega)$ denotes a nonempty diffeomorphic transformation set. The similarity between the images $X$ and $Y$ is measured by $\mathcal{L}_{sim}$, while $\mathcal{L}_{smooth}$ is a regularization function that enforces spatial smoothness of the transformation. Especially if $\bm{\phi}\left(\cdot\right)$ is a diffeomorphic mapping, thus it has an inverse mapping $\bm{\phi}^{-1}\left(\cdot\right)$ that satisfies
\begin{equation}
\bm{\phi}  \circ {\bm{\phi}^{ - 1}} = \bm{\phi} \left(\bm{\phi}^{- 1}\right) = \bm{\phi}^{ - 1}\left(\bm{\phi}\right) = \mathbb{I},
\end{equation}
where $\mathbb{I}$ stands for identity mapping.

\subsection{Image Registration via Deep Learning}
Deep learning methods use data-driven networks to align images. VoxelMorph, created by Balakrishnan et al. \cite{VoxelMorph2019}, uses the U-net architecture to achieve accurate image alignment. However, the quality of the deformation field was not optimal. To address this issue, Dalca et al. \cite{diffVoxelMorph2018} improved VoxelMorph by incorporating scaling and squaring techniques \cite{Arsigny2006}. They decomposed the deformation field into integrals of multiple velocity fields and derived the final approximate diffeomorphic deformation field through integration. This approach has been extensively adopted in later research concerning diffeomorphic registration using deep learning techniques.

The widespread use of U-net, combined with the scaling and squaring technique in medical image registration, has led to multiple frameworks for image registration. VTN in \cite{vtn2019} decomposes large displacements into smaller ones and then uses cascaded U-net networks to refine the registration process from coarse to fine levels. CycleMorph \cite{CycleMorph2020} employs cycle loss as an implicit regularization to ensure diffeomorphic registration. SYM-net \cite{sym2020} uses the U-net architecture and produces symmetric deformation fields. Kang et al. \cite{prnet2022} propose a dual-stream pyramid network that utilizes two U-Nets with shared parameters to extract features from input data at various scales. These features are then fused using a PR++ module for multi-resolution registration. Wei et al. \cite{Wei2022} incorporate an adaptive smoothing layer and an anti-folding constraint into the U-net-based registration network. Chen et al. \cite{TransMorph2022} address the challenge of limited connectivity in long-range spatial interactions within a CNN network by using TransMorph, which combines Vision Transformer with CNN. 

Inspired by quasi-conformal (QC) Teichmüller theories, Chen et al. \cite{QGChen2024} proposed a deep learning framework to learn the beltrami-coefficient for maintaining diffeomorphic registration. Using QC theories, Zhang et al. \cite{HZHANG2024} developed the topology preservation segmentation network to achieve object segmentation while preserving the topology of the image.

\subsection{Scaling and Squaring Approach}
Inspired by DARTEL \cite{Ashburner2007} and diffeomorphic Demons \cite{Vercauteren2009}, Dalca et al. implemented the scaling and squaring approach \cite{Arsigny2006} to develop a deep learning framework for diffeomorphic registration \cite{diffVoxelMorph2018}. The deformation field is represented as $\bm{\phi} = \exp (\bm{v})$, where the velocity field $\bm{v}$ is a diffeomorphic exponential flow field. Assuming a constant velocity field $\bm{v}(\bm{x})$, the relationship between the velocity and deformation fields can be expressed as follows:
\begin{equation}
\label{eq2} \frac{d\bm{\phi}\left(\bm{x},t\right)}{dt}=\bm{v}(\bm{\phi}\left(\bm{x},t\right),t),\;t\in[0,1],
\end{equation} 
where the time steps are typically set to $2^7$, and the learning algorithm with scaling and squaring approach is described as follows:
\begin{algorithm}[H]
\caption{Learning image registration with scaling and squaring approach}
\label{alg:ss}
\begin{description}
\item[S-1] Input the moving and fixed images $X$ and $Y$;
\item[S-2] Obtain the velocity field $\bm{v}(\bm{x})$ by a CNN learner, and then divide it by the time steps $2^T$ to get $\bm{v}(\bm{x})/2^T$;
\item[S-3] Calculate the deformation field per unit time step by $\bm{\phi}_{1/{2^T}} = \bm{x} + \frac{\bm{v}(\bm{x})}{2^T}$;
\item[S-4] Obtain the total deformation field $\bm{\phi}_1$ by the following recursive compound operation:
\[\begin{split}
\bm{\phi}_{1/{2^{T - 1}}} &=\bm{\phi}_{1/{2^T}} \circ\bm{\phi}_{1/{2^T}},\\
&\vdots \\
\bm{\phi}_1&=\bm{\phi}_{1/2}\circ\bm{\phi}_{1/2}.
\end{split}\]
\end{description}
\end{algorithm}

This technique guarantees a technically diffeomorphic registration. Nevertheless, employing the scaling and squaring approach requires a larger $T$ to keep $\bm{v}(\bm{x})/{2^T}$ adequately small. Unfortunately, frequent interpolation steps can cause error accumulation, thereby decreasing the accuracy of the registration and the quality of the deformation field.

\begin{figure*}
\begin{center}
\includegraphics[width=0.99\linewidth]{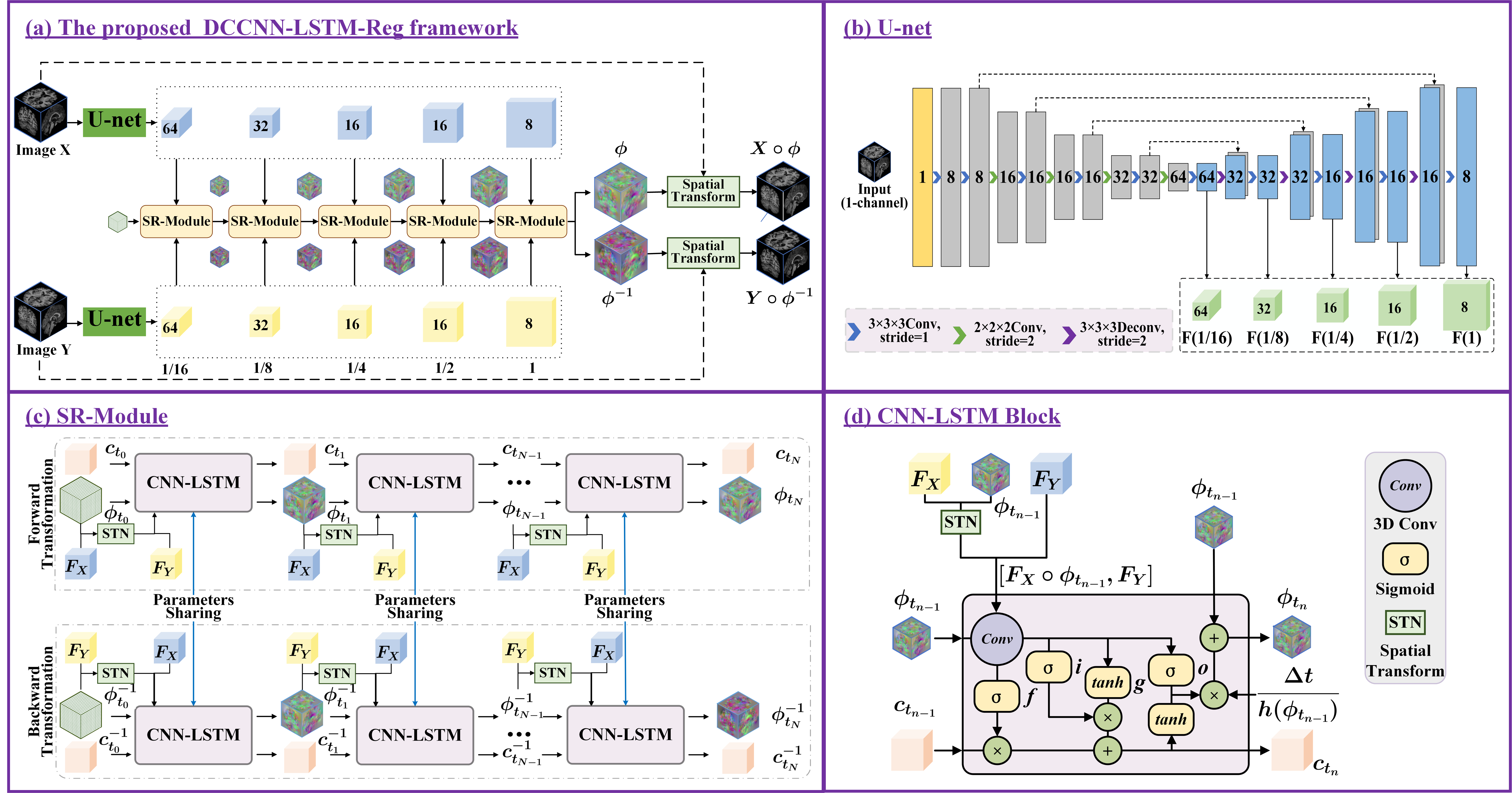}
\end{center}
\caption{The architecture of symmetric diffeomorphic Cascade CNN-LSTM image registration. 
(a) DCCNN-LSTM-Reg framework, where the symmetric configuration of the SR module in diffeomorphic image registration comprises two competing pathways that stem from the theory of one-to-one deformation. (b) The U-Net module architecture is designed to capture dual multiscale feature pyramids using shared parameters. The green block represents the multiscale features $\{\bm{F}_X^\ell\}_{\ell=1}^L$ or $\{ \bm{F}_Y^\ell\}_{\ell=1}^L$ extracted from the input images $X$ or $Y$ using the pre-trained U-Net. (c) The symmetrical diffeomorphic image registration module (SR module).  A series of $N$ CNN-LSTM blocks are connected in succession to capture the incremental field, which are then gradually integrated to produce the final pair of deformation fields. (d) CNN-LSTM block, where $[\bm{F}_X,\bm{F}_Y]$ represent the features extracted from images $X$ and $Y$ by a pre-trained Unet in Fig.\ref{fig: overview}(b), and $[\bm{F}_X\circ\bm{\phi}_{t_{n-1}},\bm{F}_Y]$ are pre-aligned by the previous deformation $\bm{\phi}_{t_{n-1}}$.
}
\label{fig: overview}
\end{figure*}

\subsection{Diffeomorphic Registration with Dynamical System}
Many traditional registration algorithms achieve the diffeomorphic transformation by integrating the velocity field $\bm{v}$ into the optimization problem. These methods guarantee that the transformation mapping $\bm{\phi}$ remains continuous and can be reversed. However, direct optimization of the velocity field presents difficulties. To address this obstacle, Zhang and Li \cite{ZHANG2022} employed a control increment $\bm{u}(\bm{\phi}\left({\bm{x},t} \right))$ to formulate $\bm{v}$ as follows:
\begin{equation}\label{eq18} 
\bm{v}(\bm{\phi}\left(\bm{x},t\right)): = \frac{\bm{u}(\bm{\phi} \left( \bm{x},t\right))}{{h(\bm{\phi} \left(\bm{x},t\right),t)}}, 
\end{equation}
where the homotopy continuation function $h(\bm{\phi}\left(\bm{x},t\right),t)>0$ incorporates the time-dependent embedding from $0$ to $1$. To achieve a diffeomorphic transformation, the incremental field $\bm{u}(\bm{\phi}\left( \bm{x},t\right))$ must also satisfy the following conditions:

\begin{equation}
\label{cic} 
\left\{ 
\begin{array}{*{20}{r}} 
\text{div}(\bm{u}(\bm{\phi}(\bm{x},t))) + \cfrac{\partial h(\bm{\phi}(\bm{x},t),t)}{\partial t} = 0,&\bm{x} \in {\Omega_{in}},\\ \bm{u}(\bm{\phi}(\bm{x},t)) = 0,&\bm{x} \in \partial \Omega . 
\end{array} \right. 
\end{equation}

The diffeomorphic image registration can be fine-tuned using the increment field $\bm{u}(\bm{\phi}\left(\bm{x},t\right))$ to ensure that it evolves smoothly over the interval $t \in [0,1]$.  In other words, the temporary deformation field $\bm{\phi}\left(\bm{x},t\right)$ of (\ref{eq2}) at each time $t$ satisfies the diffeomorphic system defined by
\begin{equation}\label{eq:zj2b}
\det \nabla \bm{\phi} (\bm{x},t)=\frac{h(\bm{x},0)}{h(\bm{\phi} (\bm{x},t),t)}>0,\; \text{ for all } t\in (0,1]. 
\end{equation}
and is also diffeomorphic. Consequently, an iterative formula can be derived to solve $\bm{\phi}\left(\bm{x},t\right)$ by implementing Euler's method as follows:
\begin{equation}\label{eqmm} 
\left\{ {\begin{array}{*{20}{l}} 
\bm{\phi}\left(\bm{x},t\right) = \bm{\phi}\left(\bm{x},t_{old}\right) + \delta t\cfrac{\bm{u}(\bm{x},{t_{old}})}{h(\bm{\phi}\left(\bm{x},t_{old}\right),t_{old})},\\ 
\bm{\phi}\left(\bm{x},0\right) = \bm{x}. 
\end{array}} \right. 
\end{equation} 

\section{Methodology}\label{section:3}
There exist many variational models for the diffeomorphic image registration task. However, the challenge is to develop a model-based learning method that ensures diffeomorphism. We introduce a new learning framework called DCCNN-LSTM-Reg, which is based on equations (\ref{eq18})-(\ref{eqmm}). Instead of directly learning the deformation field $\bm{\phi}$, our framework aims to train the time-dependent evolving control increment field $\bm{u}(\bm{\phi} \left( \bm{x},t\right))$, which governs the dynamics of the system (\ref{eqmm}). The diffeomorphic deformation field $\bm{\phi}\left(\bm{x},1\right)$ is then indirectly derived using Euler's method, where each deformation field $\bm{\phi}\left(\bm{x},t\right)$ for $t \in [0,1]$ is designed to prevent folding, thus preserving the image's topology. 

Fig. \ref{fig: overview} presents the DCCNN-LSTM-Reg framework, which includes two U-Net feature extraction sub-networks with shared parameters and a symmetric diffeomorphic registration path. A detailed introduction of the DCCNN-LSTM-Reg architecture will focus on its four principal components: 1) an U-net module for multiscale feature extraction designed to obtain dual feature pyramids; 2) a learnable control increment module (CNN-LSTM) obtained from preregistered features using homotopy continuation; 3) an SR-module designed for inversible symmetrical path registration, which integrates with cascaded CNN-LSTM blocks to learn symmetric deformation fields; and 4) different diffeomorphic losses and a similarity loss which are incorporated into our network.

\subsection{Dual Multiscale Feature Extraction}\label{subsection:unet}
Inspired by the Dual-PRNet framework \cite{prnet2022}, DCCNN-LSTM-Reg employs two dual pre-trained U-nets with shared parameters to extract feature pyramids of moving and fixed images, as shown in Fig. \ref{fig: overview}(b). The encoder comprises two 3D convolutional layers (3x3x3) with a stride of $1$, followed by four additional 3D convolutional layers (2x2x2) with a stride of 2, allowing hierarchical downsampling. Each convolutional operation is followed by a ReLU activation function. In the decoder, there are four 3D transposed convolutional layers (3x3x3) with a stride of $1$, employed for upsampling until the original image resolution is restored. The encoder and decoder are connected through skip connections, which are illustrated by dashed lines in Fig. \ref{fig: overview}(b). The output of the dual U-nets comprises multiscale features extracted from the two images on five scales, with channel numbers ranging from $\{64, 32, 16, 16, 8\}$.

Our approach differentiates itself from the alternative learning techniques such as VoxelMorph \cite{VoxelMorph2019} as we focus separately on extracting the features of the original images and discovering spatial deformations through a variational image registration mechanism. This results in more accurate and hierarchically structured deformation fields across five scales. Furthermore, our method offers greater interpretability and is more appropriate for integration with mathematical models.

\subsection{Learning Deformation Increments} \label{subsection:mathematical}
After the pyramids of the features are extracted, the proposed DCCNN-LSTM-Reg proceeds to learn the increment fields $\{u^\ell_{t_n}\}_{\ell=1,n=1}^{L,N}$ of deformation in multiple stages, and then computes the deformation fields $\{\bm{\phi}_{t_n}^{\ell}\}_{\ell=1,n=1}^{L,N}$ using formula \eqref{eqmm}, as shown in Algorithm \ref{alg:new}. 

Firstly, the dual U-net component is utilized to generate dual feature pyramids $\{(\bm{F}_X^\ell, \bm{F}_Y^\ell)\}_{\ell=1}^L$ from each pair of input images $(X, Y)$. Then, a multiscale registration process is performed, starting from the smallest (coarse) scale and progressing to the full (fine) scale. At each scale $\ell$, DCCNN-LSTM-Reg gradually learns $N$ incremental fields $\{u^\ell_{t_n}\}_{n=1}^N$ through the proposed SR module connected to $N$ CNN-LSTM units, and then refines the intermediate deformation field $\bm{\phi}_{t_n}^{\ell}$ according to Algorithm \ref{alg:new} until obtaining the final deformation field $\bm{\phi}_{1}^{\ell}:=\bm{\phi}_{t_N}^{\ell}$ at scale $\ell$. Finally,  $\bm{\phi}_{1}^{\ell}$ is upsampled by a factor of $2$ to serve as the initial deformation field $\bm{\phi}_{t_0}^{\ell+1}$ at scale $\ell+1$. This recursive process is iterated to derive $\bm{\phi}_1:=\bm{\phi}_{t_N}^{L}$ as the resolution $\ell$ increases.

To preserve the same topological structure of the image throughout each cascade, the temporary deformation field $\bm{\phi}\left(\bm{x},t\right)$ in (\ref{eqmm}) is subject to a homotopy composition denoted by $h(\bm{\phi}\left(\bm{x},t\right),t)$ \cite{ZHANG2022}. We approximate $h(\bm{\phi}\left( \bm{x},t\right),t)$ as
\begin{equation}\label{eq27}
h \approx \frac{1}{2\pi \sigma ^2}\int_\Omega h(\bm{y})\exp ( - \frac{\left\|\bm{y} -\bm{\phi}(\bm{x},t)\right\|^2}{2\sigma ^2})d\bm{y},
\end{equation}
where $h(\bm{y}):=h_o(\bm{x}) = 1$ at time $t=0$.

\begin{algorithm}
\caption{DCCNN-LSTM-Reg.}
\label{alg:new}
\begin{description}
\item[S-1] Input the moving and fixed images $X$ and $Y$, initial deformation field $\bm{\phi}_0\left(\bm{x}\right) =\bm{x}$;
\item[S-2] Extract dual pyramid features $\{(\bm{F}_X^\ell,\bm{F}_Y^\ell)\}_{\ell=1}^L$ with $L$ scales from inputs $(X,Y)$ by the dual U-net modules of our DCCNN-LSTM-Reg,  respectively;
\item[S-3] Use $N$ cascaded CNN-LSTM at each feature scale $\ell$ to gradually learn the deformation field from $t=0$ with time step-length $\frac{1}{N}$ ($n\le N$, $\ell \le L$):
\begin{description}
\item[S-3.1] Pre-align features $[\bm{F}_X \circ (\bm{\phi}^{\ell-1}_{t_{n - 1}}),\bm{F}_Y]$ and $[\bm{F}_X ,\bm{F}_Y\circ ((\bm{\phi}^{\ell-1}_{t_{n - 1}})^{-1})]$;
\item[S-3.2] Use cascade CNN-LSTM to learn the control incremental fields $u_{t_n}^\ell$ and $(u_{t_n}^\ell)^{-1}$ at time $t_n=\frac{n}{N}$;
\item[S-3.3] Iteratively update the current deformation field by 
\[\begin{split}
\bm{\phi}_{t_n}^\ell&=\bm{\phi}_{t_{n-1}}^\ell+\delta t \cdot \frac{u_{t_n}^\ell}{h(\bm{\phi}_{t_{n - 1}}^\ell)},\\
(\bm{\phi}_{t_n}^\ell)^{-1}&=(\bm{\phi}_{t_{n-1}}^\ell)^{-1}+\delta t \cdot \frac{(u_{t_n}^\ell)^{-1}}{h((\bm{\phi}_{t_{n - 1}}^\ell)^{-1})};
\end{split}
\]
\end{description}
\item[S-4] Output final solutions ${\bm{\phi}_1}\left(\bm{x} \right)$ and $(\bm{\phi}_1\left(\bm{x} \right))^{-1}$.
\end{description}
\end{algorithm}

Unlike the scaling and squaring approach, our registration technique integrates control increments directly into the deformation field, thereby minimizing the error amplification that results from repeated interpolations. Additionally, the use of a variable velocity field enhances adaptability, enabling simple and direct corrections to the deformation field.

\subsection{SR-Module for Symmetric Multiscale Registration}  
The SR-module we propose, illustrated in Fig. \ref{fig: overview}(c), is designed to achieve symmetric deformations for moving and fixed images. It focuses on the analysis of sequential increment fields $u_{t_n}^\ell$ across five scales ($\ell=1,\dots,5$). Previous work \cite{vtn2019} suggested cascaded U-net networks, which can be resource intensive and prone to overfitting. Moreover, interpolation of U-net may compromise the accuracy of the registration. To address these issues, we propose a symmetric cascade CNN-LSTM increment learning module simplified as an SR-module.

Inspired by the Conv-LSTM mechanism introduced in \cite{ConvLSTM2015}, a CNN-LSTM block, illustrated in Fig. \ref{fig: overview}(d), is designed to capture spatial deformation corrections between images for multiple time steps. Initially, CNN-LSTM takes into account the features $(\bm{F}_X,\bm{F}_Y)$ along with an identical deformation field as input. As the cascade progresses, CNN-LSTM handles the pre-aligned features $[\bm{F}_X \circ (\bm{\phi}^\ell_{t_{n - 1}}),\bm{F}_Y]$ and the deformation field $\bm{\phi}^\ell_{t_{n - 1}}$ from the previous step. By connecting CNN-LSTM blocks across multiple cascades, DCCNN-LSTM-Reg can integrate local increment at each time $t_n$ and continuously improve global deformation fields, thereby facilitating a gradual multiscale registration from coarse to fine levels. The operation of CNN-LSTM at cascade $t_n$ can be defined as:
\begin{equation} 
\begin{split}%{c} 
f_{t_n},i_{t_n},g_{t_n},o_{t_n} &= Conv(\bm{F}_X \circ (\bm{\phi}^\ell_{t_{n - 1}}),\bm{F}_Y,\bm{\phi}^\ell_{t_{n - 1}});\\ 
c_{t_n} &= \sigma (f_{t_n}) \cdot c_{t_{n - 1}} + \sigma (i_{t_n}) \cdot \tanh (g_{t_n});\\ 
u^\ell_{t_n} &= \sigma (o_{t_n}) \cdot \tanh \left( c_{t_n} \right);\\ 
\bm{\phi}^\ell_{t_n} &= \bm{\phi}^\ell_{t_{n - 1}} + \Delta t \cdot \frac{u^\ell_{t_n}}{h(\bm{\phi}^\ell_{t_{n - 1}} )}. \nonumber
\end{split} 
\end{equation}

The combination between $\bm{F}_X \circ (\bm{\phi^\ell}_{t_{n - 1}})$, $\bm{F}_Y$, and $\bm{\phi}^\ell_{t_{n - 1}}$ relies on the channel dimension, which acts as an input to the convolutional layer. Consequently, the resulting output is divided into four intermediary features that include the input feature $i$, forgotten feature $f$, output feature $o$, and update feature $g$, each of which shares the same dimension as the deformation field. The updated memory feature $c_{t_n}$ preserves information from the current and all preceding cascades. Through three gating mechanisms, CNN-LSTM produces the increment field $u^\ell_{t_n}$ at cascade $t_n$, which is adjusted proportionally by $\frac{{\Delta t}}{{h(\bm{\phi}^\ell_{t_{n - 1}} )}}$ and then combined with the deformation field $\bm{\phi}^\ell_{t_{n - 1}}$ from the previous cascade to obtain the deformation field $\bm{\phi}^\ell_{t_{n}}$. By reversing the order of the feature volumes $\bm{F}_X$ and $\bm{F}_Y\circ((\bm{\phi}_1^\ell)^{-1})$ and using shared weights, we establish the symmetrical registration path. 

The SR model integrates a fusion of CNN-LSTM blocks to handle forward and backward deformation fields for multiscale registration. The structure of the SR-module is illustrated in Fig. \ref{fig: overview}(c), with each path consisting of $N$ CNN-LSTM blocks. This strategy helps reduce the number of parameters, prevent overfitting, and facilitate training.

\subsection{Loss Function}
Our loss system for DCCNN-LSTM-Reg incorporates five elements: similarity loss, smoothness loss, Jacobian loss, cycle consistency loss, and control incremental constraint. The primary aim of similarity loss is to enhance the correlation between images. Conversely, the smoothness loss, Jacobian loss, cycle consistency loss, and control incremental constraint work together to maintain the smoothness and diffeomorphism of the registration grid. 

\subsubsection{Similarity Loss}
We employ the Normalized Cross-Correlation (NCC) \cite{ncc2005} to measure similarity. Our method takes into account both the forward and backward registration steps, as well as the registration outputs at multiple scales ($1\leq\ell\leq L$). The similarity loss is formally expressed as: \begin{equation}
\begin{split}
\mathcal{L}_{sim} = -\sum\limits_\ell^L\lambda _1^{\ell}\big(&\text{NCC}(X\circ (\mathcal{P}(\bm{\phi}_1^\ell)),Y)\\&+\text{NCC}(X,Y \circ (\mathcal{P}((\bm{\phi}^\ell)_1^{-1})))\big),\nonumber
\end{split}
\end{equation} 
in which $X$ and $Y$ represent the moving and fixed images, respectively. $L$ stands for the total number of scales, $\bm{\phi}_1^\ell$ and $(\bm{\phi}_1^\ell)^{ - 1}$ correspond to the deformations generated by DCCNN-LSTM-Reg in both forward and reverse directions across multiple scales. $\lambda_1^{\ell}$ is a parameter that is used to weigh the importance of similarity loss on different scales. The term $\mathcal{P}$ refers to the trilinear upsampling operator, where the deformation fields on scales $\ell=5, 4, 3, 2$ are upsampled to the full scale for loss computation.

\subsubsection{Jacobian Loss}
In the DCCNN-LSTM-Reg framework, we incorporate a Jacobian loss $\mathcal{L}_{Jdet}$ to ensure the maintenance of the diffeomorphism characteristic of the deformation field. This loss is applied to both the forward deformation field $\bm{\phi}_1^\ell$ and its inverse $(\bm{\phi}_1^\ell)^{-1}$ at each level. It is determined by evaluating the negative values of the Jacobian determinant for each point on the registration grid, employing the Rectified Linear Unit (ReLU) activation function: \[\mathcal{L}_{Jdet}=\sum\limits_\ell^L\sum\limits_{\bm{x}\in\Omega }\left(\text{Relu}(-J_{\bm{\phi}_1^\ell}(\bm{x}) )+\text{Relu}(-J_{(\bm{\phi}_1^\ell)^{-1}}(\bm{x}))\right).\] 

\subsubsection{Smooth Loss}
To ensure that the deformation fields remain smooth across all levels, DCCNN-LSTM-Reg incorporates a $\ell_2$ regularization term into the gradient fields of the deformations. The smoothness loss can be defined as: 
\[\mathcal{L}_{reg} =\sum\limits_\ell^L\sum\limits_{\bm{x}\in\Omega}\left(\parallel\nabla\bm{\phi}_1^\ell\left(\bm{x}\right)\parallel _2^2+\parallel \nabla(\bm{\phi}^\ell)_1^{-1}\left(\bm{x}\right)\parallel_2^2\right),\] 
which ensures the smoothness of the forward and backward deformation fields.

\subsubsection{Cycle Consistency Loss}
Since DCCNN-LSTM-Reg performs a symmetric registration path and generates a pair of two-way deformation fields simultaneously, it enables the creation of a cycle consistency loss based on these fields. The loss of cycle consistency can be expressed as: 
\[\mathcal{L}_{cycle} = -\text{NCC}(X,X(\bm{\phi}_1\circ {\bm{\phi}_1^{ - 1}}))-\text{NCC}(Y,Y(\bm{\phi}_1^{-1}\circ\bm{\phi}_1)).\]

\subsubsection{Control Incremental Constraint}
According to the algorithm proposed in \cite{ZHANG2022}, the control increment field $\bm{u}(\bm{\phi} \left({\bm{x},t} \right))$ in the DCCNN-LSTM-Reg framework should satisfy equation \eqref{cic}. To ensure consistency with the diffeomorphic theory, an additional constraint is added to the total loss function, which is expressed as: 
\[
\begin{split}
\mathcal{L}_{cic} =\sum\limits_l^L\Big\{& \left|\text{div}(\bm{u}(\bm{\phi}^\ell)) + \frac{\partial h(\bm{\phi}^\ell,t)}{\partial t}\right|\\
 &+\left|\text{div}(\bm{u}((\bm{\phi}^\ell)^{ - 1}))+\frac{\partial h((\bm{\phi}^\ell)^{-1},t)}{\partial t}\right|\Big\}.
\end{split}\]

\subsubsection{Total Loss}
The total training loss of our DCCNN-LSTM-Reg can be expressed as: 
\[\mathcal{L}(X,Y)=\mathcal{L}_{sim}+\lambda _2\mathcal{L}_{Jdet}+ \lambda_3\mathcal{L}_{reg}+\lambda_4\mathcal{L}_{cycle}+\lambda_5\mathcal{L}_{\rm{cic}},\] 
where $\lambda^\ell_1$, $\lambda_2$, $\lambda_3$, $\lambda_4$ and $\lambda_5$ are the weights of the similarity loss on each scale $\ell$, Jacobian loss, smoothness loss, cycle consistency loss, and control incremental constraint loss, respectively.

\begin{table*}
\centering  
\renewcommand{\arraystretch}{1.2}% Tighter
\newcolumntype{C}[1]{>{\centering\arraybackslash}p{#1}}
\setlength\tabcolsep{0.5pt}
\caption{Quantitative evaluation of three Brain MRI registration datasets (OASIS-V1, IXI, Mindboggle101) .}\label{tab: Objective_Assessment}
  %\begin{tabular}{cccccccccc}
\scriptsize
% \footnotesize
\begin{tabular*}{\hsize}{@{}@{\extracolsep{\fill}}clcccccccc@{}}
\toprule[1.5pt]
\textbf{Dataset} &  \textbf{Metrics} & \textbf{Affine} & \textbf{SyN } & \textbf{VoxelMorph} & \textbf{VoxelMorph-diff} & \textbf{TransMorph} & \textbf{TransMorph-diff} & \textbf{SYM-net} & \textbf{Proposed} \\
% ~\\
\midrule[0.8pt]
\multirow{4}{*}{\rotatebox{90}{\textbf{Inter-patient}}} 
& \textbf{DSC}        & $0.601\pm0.063$ & $0.759\pm0.031$ & $0.768\pm0.034$ & $0.728\pm0.047$&$0.784\pm0.048$ & $0.749\pm0.038$& \underline{$0.791\pm0.027$}& \bm{$0.809\pm0.018$}\\
& \textbf{HD}         & $3.584\pm0.852$ & $2.205\pm0.482$ & $2.385\pm0.619$ & $2.559\pm0.650$&\underline{$2.037\pm0.491$} & $2.261\pm0.456$& $2.100\pm0.504$& \bm{$1.906\pm0.337$}\\
& \textbf{SSIM}       & $0.678\pm0.018$ & $0.819\pm0.015$ & $0.909\pm0.009$ & $0.798\pm0.016$&\underline{$0.924\pm0.009$} & $0.887\pm0.011$& $0.920\pm0.008$& \bm{$0.932\pm0.007$}\\
&$\%\left|J_\phi\right|\le0$ & - & \bm{${\rm{<0}}{\rm{.0001}}$} & $1.3531\pm0.1978$ & \bm{${\rm{ < 0}}{\rm{.0001}}$}&$0.6202\pm0.1315$& \bm{${\rm{ < 0}}{\rm{.0001}}$}& $0.0010\pm0.0004$&\bm{${\rm{<0}}{\rm{.0001}}$} \\
    ~\\
    \multirow{4}{*}{\rotatebox{90}{\textbf{Patient-to-atlas}}} 
& \textbf{DSC} & $0.406\pm0.035$ & $0.659\pm0.038$ & $0.729\pm0.026$ & $0.705\pm0.027$&$0.746\pm0.021$ & $0.721\pm0.031$& \underline{$0.749\pm0.020$}&\bm{$0.751\pm0.018$}\\
& \textbf{HD} & $6.477\pm0.669$ & $4.501\pm0.781$ & $3.691\pm0.670$ & $3.274\pm0.495$&$3.033\pm0.422$ & $3.214\pm0.505$& \bm{$3.022\pm0.475$}&\underline{$3.026\pm0.404$}\\
& \textbf{SSIM} & $0.621\pm0.013$ & $0.796\pm0.020$ & \underline{$0.881\pm0.014$} & $0.752\pm0.017$&\bm{$0.891\pm0.018$} & $0.745\pm0.019$& $0.876\pm0.014$& $0.862\pm0.012$\\
&$\%\left|J_\phi\right|\le0$ & - & \bm{${\rm{<0}}{\rm{.0001}}$} & $1.9460\pm0.2539$ & \bm{${\rm{ < 0}}{\rm{.0001}}$}&$1.5014\pm0.1152$& \bm{${\rm{ < 0}}{\rm{.0001}}$}& $0.0005\pm0.0003$&\bm{${\rm{ < 0}}{\rm{.0001}}$} \\
    ~\\
    \multirow{4}{*}{\rotatebox{90}{\textbf{Few-shot }}} 
& \textbf{DSC}        & $0.393\pm0.020$ & $0.550\pm0.010$ & \underline{$0.600\pm0.015$} & $0.534\pm0.013$&$0.579\pm0.029$ & $0.521\pm0.011$& $0.574\pm0.015$& \bm{$0.618\pm0.010$}\\
& \textbf{HD} & $6.680\pm0.513$ &$\underline{5.598\pm0.339}$& $5.764\pm0.415$ & $5.833\pm0.412$&$5.940\pm0.409$ & $5.706\pm0.364$& $5.935\pm0.442$& \bm{$5.475\pm0.290$}\\
& \textbf{SSIM} & $0.653\pm0.013$ & $0.806\pm0.008$ & $0.927\pm0.006$ & $0.806\pm0.011$&\bm{$0.936\pm0.012$} & $0.783\pm0.010$& $0.898\pm0.006$& \underline{$0.928\pm0.006$}\\
&$\%\left|J_\phi\right|\le0$ & - & \bm{${\rm{<0}}{\rm{.0001}}$} & $1.7006\pm0.2220$ & $0.0002\pm0.0002$&$1.9144\pm0.2418$& $0.0002\pm0.0002$& $0.0009\pm0.0005$&\bm{${\rm{<0}}{\rm{.0001}}$} \\
\bottomrule[1.5pt]
    \end{tabular*}
\end{table*}

\begin{table*}
\centering  
\renewcommand{\arraystretch}{0.95}% Tighter
\newcolumntype{C}[1]{>{\centering\arraybackslash}p{#1}}
\caption{Quantitative evaluations of SYM-net and DCCNN-LSTM-Reg on part of the images on OASIS.}\label{tab: DCCNN-LSTM-Reg-part}
\begin{tabular*}{\hsize}{@{}@{\extracolsep{\fill}}llccccccccc@{}}
\toprule[1.5pt]
 & \textbf{Affine} & \multicolumn{2}{c}{\textbf{VoxelMorph-diff}} & \multicolumn{2}{c}{\textbf{TransMorph-diff}} & \multicolumn{2}{c}{\textbf{SYM-net}} & \multicolumn{2}{c}{\textbf{Proposed}} \\
%\midrule
\cmidrule(r){3-4} \cmidrule(r){5-6} \cmidrule(r){7-8} \cmidrule(r){9-10} 
 & \textbf{DSC} & \textbf{DSC} & $ \bm{\left|J_\phi\right|\le0} $ & \textbf{DSC} & $ \bm{\left|J_\phi\right|\le0} $ & \textbf{DSC} & $ \bm{\left|J_\phi\right|\le0} $ & \textbf{DSC} & $\bm{\left|J_\phi\right|\le0}$\\
\midrule[0.8pt]
\textbf{Image 1} & $0.566$ &$0.720$ & $6$ &$0.692$ &$0$ & $0.789$ & $13$ & \bm{$0.812$} & $0$ \\
%\midrule
\textbf{Image 2} & $0.590$ &$0.745$ & $1$ &$0.735$ &$0$ & $0.805$ & $50$ & \bm{$0.823$} & $0$ \\
%\midrule
\textbf{Image 3} & $0.624$ &$0.761$ & $0$ &$0.799$ &$0$ & $0.812$ & $55$ & \bm{$0.832$} & $0$ \\
%\midrule
\textbf{Image 4} & $0.573$ &$0.725$ & $14$&$0.745$ &$0$ & $0.788$ & $51$ & \bm{$0.808$} & $0$ \\
%\midrule
\textbf{Image 5} & $0.553$ &$0.718$ & $0$&$0.697$ &$0$ & $0.792$ & $52$& \bm{$0.794$} & $0$ \\
%\midrule
\textbf{Image 6} & $0.610$ &$0.748$ & $13$&$0.741$ &$0$ & $0.786$ & $56$ & \bm{$0.808$}  & $0$ \\
%\midrule
\textbf{Image 7} & $0.640$ &$0.759$ & $10$&$0.736$ &$0$ & $0.807$ & $35$ & \bm{$0.821$}  & $0$ \\
%\midrule
\textbf{Image 8} & $0.616$ &$0.763$ & $3$&$0.782$ &$0$ & $0.810$ & $33$ & \bm{$0.826$}  & $0$ \\
\bottomrule[1.5pt]
\end{tabular*}
\end{table*}

\section{Experiment}\label{section:4}
\subsection{Experimental Settings}
\subsubsection{Inter-patient Brain MRI Registration}
This study involves conducting experiments on aligning brain MRI scans of different patients using the OASIS-v1 dataset \cite{oasis2007, Hoopes2022}. The dataset comprises 414 T1-weighted MRI images and their segmentation labels. The preprocessing was performed with Freesurfer \cite{freesurfer2012}, covering tasks such as motion correction, skull removal, affine transformations, and segmentation of the subcortical structure. The images were resized from dimensions $160\times192\times224$ to $160\times160\times192$. The dataset division included 255 images for the training set, nine for validation, and 150 for testing. During training, image pairs for alignment were randomly selected, resulting in 64,770 pairs. For validation, one image was fixed and the remaining eight were treated as moving images, forming eight pairs of validation images. In the testing, five images from the test set were chosen, with one randomly set as fixed image per iteration. Moving images were selected from the remaining 145, generating 725 test image pairs. The segmentation labels covered 35 anatomical structures to assess the accuracy of the registration.

\subsubsection{Patient-to-atlas Brain MRI Registration}
The dataset utilized in this study, provided by Chen et al. \cite{TransMorph2022}, was used to align brain MRI scans between patient data and an atlas. It includes 576 T1-weighted MRI brain scans in addition to an atlas image. The moving images originated from the IXI dataset, while the fixed images were obtained from the research by Kim et al. \cite{CycleMorph2020}. This dataset was divided into training, validation, and testing subsets with ratios of 403:58:115 (7:1:2). Each image was resized to dimensions of $160\times160\times192$. To assess registration accuracy, segmentation was performed in 30 different anatomical regions.

\begin{figure*}[t!]
\centering
\includegraphics[width=\linewidth]{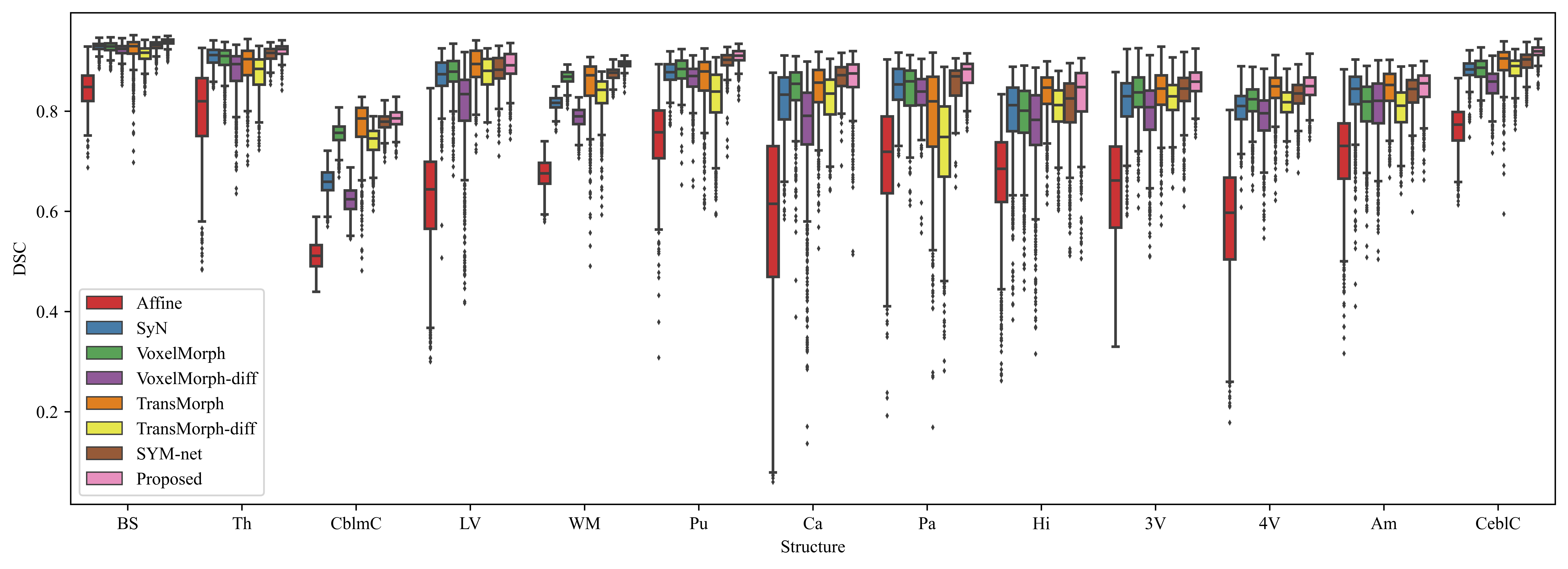}
\caption{Comparison of DSC scores for each anatomical region between state-of-the-art methodologies and our proposed approach. To enhance clarity, the left and right brain hemispheres were combined into a single region. The structures analyzed included the brain stem (BS), thalamus (Th), cerebellar cortex (CblmC), lateral ventricle (LV), cerebellar white matter (WM), putamen (Pu), caudate (Ca), pallidum (Pa), hippocampus (Hi), 3rd ventricle (3V), 4th ventricle (4V), amygdala (Am), CSF (CSF), and cerebral cortex (CeblC).}
\label{fig: DCCNN-LSTM-Reg-Dice}
\end{figure*}

\begin{table*}
\centering  
\renewcommand{\arraystretch}{0.95}% Tighter
\newcolumntype{C}[1]{>{\centering\arraybackslash}p{#1}}
\caption{Quantitative evaluations of SYM-net and DCCNN-LSTM-Reg on OASIS and IXI datasets for symmetric registration results.}\label{tab: DCCNN-LSTM-Reg-dual-OASIS}
\begin{tabular*}{\hsize}{@{}@{\extracolsep{\fill}}llcccc@{}}
\toprule[1.5pt]
\textbf{Dataset} & \textbf{Model} & \multicolumn{2}{c}{\textbf{Forward Registration} ($\bm{X\rightarrow Y}$)}& \multicolumn{2}{c}{\textbf{Backward Registration} ($\bm{X\leftarrow Y}$)} \\
%\midrule
\cmidrule(r){3-4} \cmidrule(r){5-6}
 & & \textbf{DSC} & $\bm{\%\left|J_\phi\right|\le0} $ & \textbf{DSC} & $\bm{\%\left|J_\phi\right|\le0}$\\
\midrule[0.8pt]
& Affine & $0.601\pm0.063$ & - & $0.601\pm0.063$ & - \\
%\midrule
\textbf{OASIS} & SYM-net & $0.791\pm0.027$ & $0.0010\pm0.0004$ & $0.792\pm0.028$ & $0.0008\pm0.0003$ \\
%\midrule
& Proposed & \bm{$0.809\pm0.018$} & \bm{${\rm{<0}}{\rm{.0001}}$} & \bm{$0.809\pm0.019$} & \bm{${\rm{<0}}{\rm{.0001}}$} \\
\\
& Affine & $0.406\pm0.035$ & - & $0.406\pm0.035$ & - \\
%\midrule
\textbf{IXI} & SYM-net & $0.749\pm0.020$ & $0.0005\pm0.0003$& \bm{$0.732\pm0.024$} & $0.0016\pm0.0006$ \\
%\midrule
& Proposed & \bm{$0.751\pm0.018$}  & \bm{${\rm{<0}}{\rm{.0001}}$} & \bm{$0.732\pm0.019$}  & \bm{${\rm{<0}}{\rm{.0001}}$} \\
\bottomrule[1.5pt]
\end{tabular*}
\end{table*}

\subsubsection{Few-Shot Dataset MRI Registration}
In our Few-Shot MRI Registration study, we used the Mindboggle101 dataset \cite{mindboggle101}, focusing on the NKI-RS-22, NKI-TRT-20, and OASIS-TRT-20 subsets, which together provided 62 T1-weighted brain MRI scans. Originally aligned in the MNI152 space with a resolution of $182\times218\times182$, these images were subsequently resized to dimensions of $160\times192\times160$. The dataset was divided into a training set of 50 images and a testing set of 12 images. During training, we randomly selected pairs from the training set to generate 2,450 pairs of training images. In the testing phase, one image was served as the fixed reference, while the other 11 were used as moving images, forming 11 test image pairs.

\subsubsection{Comparison Methodology}
Our proposed DCCNN-LSTM-Reg model is evaluated against a traditional variational method and five deep learning methods. The selected methods include SyN \cite{syn2008}, VoxelMorph \cite{VoxelMorph2019}, VoxelMorph-Diff \cite{diffVoxelMorph2018}, TransMorph \cite{TransMorph2022}, TransMorph-Diff \cite{TransMorph2022}, and SYM-net \cite{sym2020}. In the comparative analysis, we apply the optimal parameter configurations as specified in the original publications.

\subsubsection{Evaluation Metrics}
The performance of the registration was assessed by examining the anatomical features of the aligned images using the Dice similarity coefficient (DSC) and the Hausdorff distance (HD). A quantitative evaluation involved comparing the mean and standard deviation of DSC and HD for the designated anatomical features among all patients. The structural similarity index (SSIM) was employed to evaluate the similarity between the fixed image and the registered image. Furthermore, the percentage of nonpositive values in the determinant of the Jacobian matrix in the deformation field, denoted $\% |J_{\bm{\phi}}| \le 0$, was used to measure the folding ratios of the registration field.

\subsubsection{Implementation}
The Python language (version 3.11.5) and the PyTorch deep learning framework (version 2.1.2) were utilized on a Linux OS (Ubuntu 22.04.1 LTS). The hardware setup included a 12th-gen Intel (R) Core (TM) i7-12700F CPU and a single NVIDIA GeForce RTX 4090 GPU. For DCCNN-LSTM-Reg, the learning rate was set to ${10}^{-4}$ and a batch size of 1 was used. The weights of the loss function were fixed as $\lambda_1=0.8$, $\lambda_2=1 \times {10^5}$, $\lambda_3=1$, $\lambda_4=0.1$, and $\lambda_5=0.1$.

\subsection{Comparisons baselines}
In this section, we evaluate the effectiveness of our proposed approach by comparing it to the baseline methods on three datasets. This evaluation involves both qualitative and quantitative analyses.

\subsubsection{Objective Assessment}
Firstly, the evaluation of the effectiveness of the DCCNN-LSTM-Reg registration model on three brain MRI datasets is carried out by analyzing the accuracy of the deformation field and detecting any folding. Table \ref{tab: Objective_Assessment} presents the evaluation results of our proposed method alongside several comparison techniques for three medical image registration tasks. The metrics evaluated include DSC, HD, SSIM and $\%|J_{\bm{\phi}}|\le 0$. The best results are emphasized in bold, while the second-best are underlined.

Table \ref{tab: Objective_Assessment} demonstrates that our proposed approach ranks within the top two for 11 out of 12 evaluation metrics, where it achieves the first position in nine of these metrics. In particular, our technique achieved the highest DSC metrics in all three datasets. In comparison with SyN and two other diffeomorphic deep learning models, our approach maintained a comparable deformation field folding rate (${\rm{<0}}{\rm{.0001}}$) but significantly outperformed them in registration accuracy, with a maximum discrepancy exceeding $10\%$. As depicted in Fig. \ref{fig: DCCNN-LSTM-Reg-Dice}, DCCNN-LSTM-Reg demonstrated higher precision and fewer irregularities for more than half of the anatomical structures compared to the other methods.

Table \ref{tab: DCCNN-LSTM-Reg-part} presents the results of the quantitative analysis for four models applied to eight image pairs from the OASIS dataset. Our proposed method achieves the highest DSC score and yields a deformation field without fold points. In contrast, both the VoxelMorph and the SYM-Net exhibit varying degrees of folding. Although TransMorph does not present any fold points, its DSC score is considerably lower than that of our proposed approach.

The results of the objective assessment reveal the efficiency of the DCCNN-LSTM-Reg model, which show that our framework achieves superior registration performance and outperforms several networks that use scaling and squaring techniques to handle deformation folds.

\begin{table}
\centering  
\renewcommand{\arraystretch}{0.95}% Tighter
\newcolumntype{C}[1]{>{\centering\arraybackslash}p{#1}}
\caption{Quantitative assessments of ablation studies conducted on the OASIS dataset using S-DCCNN-LSTM-Reg with and without various modules.}\label{tab: DCCNN-LSTM-Reg-ablation}
%\begin{tabular}{C{3cm} C{2cm} C{2cm}}
\begin{tabular*}{\hsize}{@{}@{\extracolsep{\fill}}lcc@{}}
\toprule[1.5pt]
\textbf{Variants} &  \textbf{DSC} & $ \bm{\%\left|J_\phi\right|\le0} $\\
\midrule[0.8pt]
\textbf{S-DCCNN-LSTM-Reg}\\
%\midrule
\quad (a) r/ Conv-GRU & $0.704\pm0.067$  & $0.0012\pm0.0003$ \\
%\midrule
\quad (b) r/ Resnet & $0.782\pm0.028$  & ${\rm{<0}}{\rm{.0001}}$ \\
%\midrule
\quad (c) r/ TransMorph & $0.787\pm0.022$  & ${\rm{<0}}{\rm{.0001}}$ \\
%\midrule
\quad (d) w/o Cycle Loss & $0.793\pm0.030$  & $0.0019\pm0.0007$ \\
%\midrule
\quad (e) w/o Smooth Loss & $0.775\pm0.034$  & $0.0059\pm0.0011$ \\
%\midrule
\quad (f) w/o Jacobian Loss & $0.789\pm0.033$  & $1.6990\pm0.2793$ \\
%\midrule
\quad (g) w/o symmetrical path & $0.791\pm0.032$  & $0.0020\pm0.0007$ \\
%\midrule
\quad (h) w/ deformed features & \bm{$0.809\pm0.024$}  & $0.0005\pm0.0003$ \\
%\midrule
\quad (i) Baseline& \bm{$0.794\pm0.027$}  & \bm{$0.0002\pm0.0001$} \\
%\midrule
\textbf{DCCNN-LSTM-Reg}\\
%\midrule
\quad (j) Baseline (w/$\mathcal{L}_{cic}$) & \bm{$0.809\pm0.018$}  & ${\rm{<0}}{\rm{.0001}}$ \\
\bottomrule[1.5pt]
\end{tabular*}
\end{table}

\begin{table*}
\centering  
\renewcommand{\arraystretch}{0.95}% Tighter
\newcolumntype{C}[1]{>{\centering\arraybackslash}p{#1}}
\caption{Quantitative evaluations of the complexity ablation experiments on OASIS dataset.}\label{tab: DCCNN-LSTM-Reg-OASIS-N}
%\begin{tabular}{cccccc}
 \begin{tabular*}{\hsize}{@{}@{\extracolsep{\fill}}lccccc@{}}
\toprule[1.5pt]
\textbf{Cascade} $\bm{\ell \times N}$ & \textbf{DSC} & \textbf{HD} & \textbf{SSIM} & $ \bm{\%\left|J_\phi\right|\le0} $ & \textbf{Parameter amount} \\
\midrule
$5\times1$&$0.764\pm0.037$&$2.274\pm0.535$&$0.837\pm0.015$&$0.0002\pm0.0001$& 767224 \\
$5\times2$&$0.783\pm0.030$&$2.131\pm0.470$&$0.872\pm0.013$&\bm{$0.0001\pm0.0001$}& 870640 \\
$5\times3$&$0.789\pm0.029$&$2.101\pm0.463$&$0.887\pm0.012$&\bm{$0.0001\pm0.0001$}& 974056 \\
$5\times4$&$0.794\pm0.027$&$2.050\pm0.440$&$0.894\pm0.012$&$0.0002\pm0.0001$& 1077472 \\
$5\times5$&$0.797\pm0.029$&\bm{$2.033\pm0.450$}&$0.904\pm0.012$&$0.0002\pm0.0001$ & 1180888\\
$5\times6$&\bm{$0.798\pm0.029$}&$2.037\pm0.462$&\bm{$0.907\pm0.011$}&\bm{$0.0001\pm0.0001$}& 1284304 \\
\bottomrule[1.5pt]
\end{tabular*}
\end{table*}

\begin{figure*}[t!]
  \centering  \includegraphics[width=\linewidth]{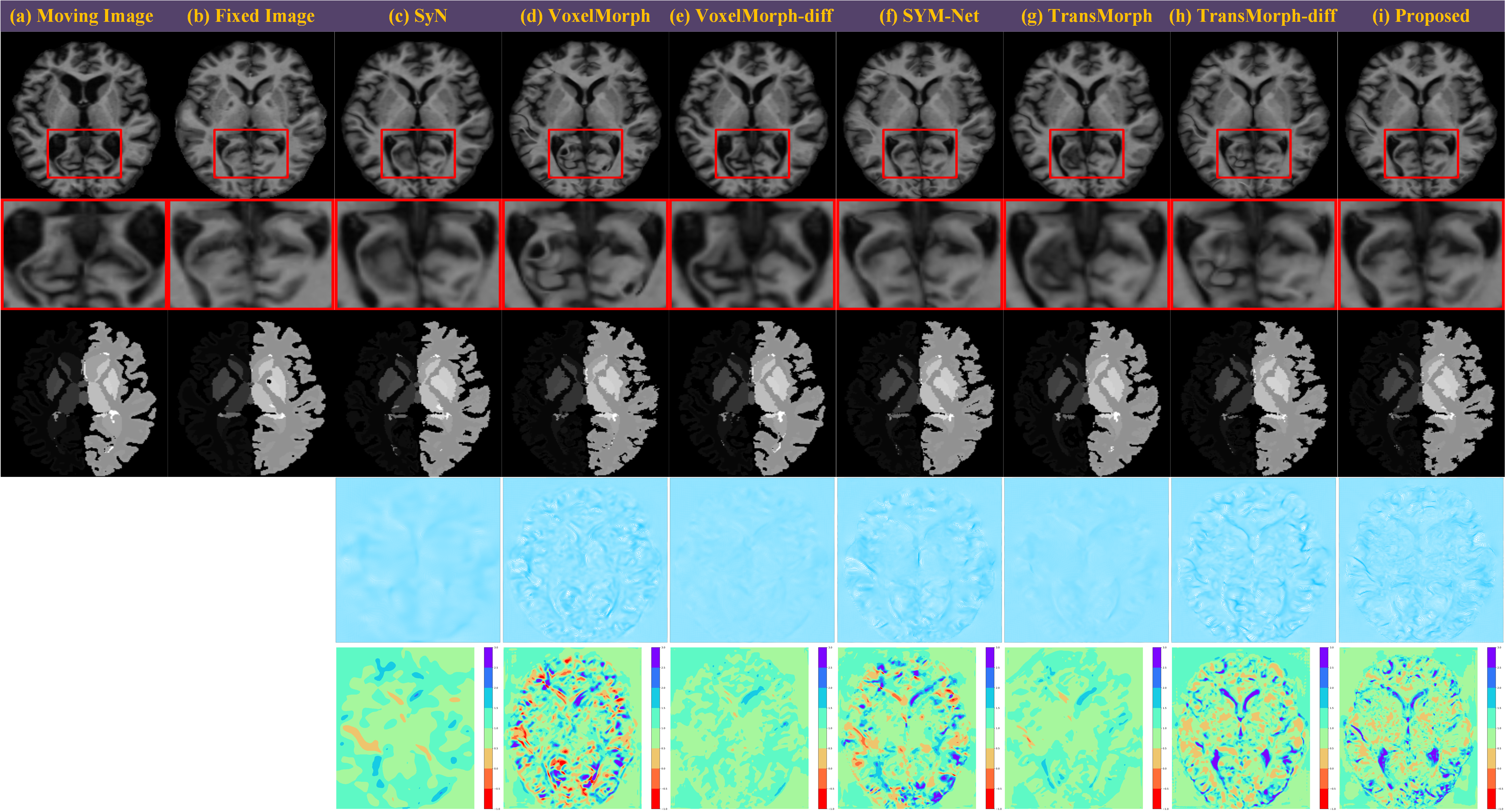}
  \vspace{-2pt}
  \caption{Comparisons with different registration methods for one pair of MRI images. From top to bottom: original images and registered images, local zoom-in of the original images, segmented images, deformation fields, heat maps of Jacobian determinants.}
  \label{fig: OASISshow1}
\end{figure*}

\subsubsection{Visualization Results}
The OASIS dataset is employed to visually assess the performance of seven different methods on both the original image and the anatomical structure. As depicted in Fig. \ref{fig: OASISshow1}, DCCNN-LSTM-Reg (Fig. \ref{fig: OASISshow1}(i)) demonstrates the highest accuracy and deformation grid quality. While the Jacobian determinant heat map indicates significant folding during registration with VoxelMorph (Fig. \ref{fig: OASISshow1}(d)) and TranMorph (Fig. \ref{fig: OASISshow1}(g)). In contrast, VoxelMorph-diff (Fig. \ref{fig: OASISshow1}(e)) and TransMorph-diff (Fig. \ref{fig: OASISshow1}(h)) show reduced grid folding at the expense of lower accuracy. SYM-net (Fig. \ref{fig: OASISshow1}(f)) and DCCNN-LSTM-Reg (Fig. \ref{fig: OASISshow1}(i)) methods of symmetric registration achieve similar accuracy, yet DCCNN-LSTM-Reg (Fig. \ref{fig: OASISshow1}(i)) results in fewer grid folds. Moreover, SYM-net (Fig. \ref{fig: OASISshow1}(f)) lacks the precision of DCCNN-LSTM-Reg (Fig. \ref{fig: OASISshow1}(i)) in the texture details of the registered image, indicating that the multiscale cascade network enables DCCNN-LSTM-Reg (Fig. \ref{fig: OASISshow1}(i)) to handle deformations of various sizes more efficiently.

\subsubsection{Symmetric Registration Evaluation}
DCCNN-LSTM-Reg was compared to SYM-net using the OASIS and IXI datasets (see Table \ref{tab: DCCNN-LSTM-Reg-dual-OASIS}), and it outperformed SYM-net in both forward and reverse registration tasks. On the OASIS dataset, DCCNN-LSTM-Reg achieved a superior DSC score of 0.809 for both registration directions. Furthermore, $\% |J_{\bm{\phi}}| \le0$ was significantly lower for DCCNN-LSTM-Reg compared to SYM-net. On the IXI dataset, DCCNN-LSTM-Reg also outperformed SYM-net in both forward and reverse registrations, where SYM-net demonstrated a three-time increase in value $\% |J_{\bm{\phi}}| \le 0$ when assessing backward registration compared to forward registration, whereas DCCNN-LSTM-Reg maintained a consistently low level without any increase.

\begin{figure}[t!]
  \centering
  \includegraphics[width=\linewidth]{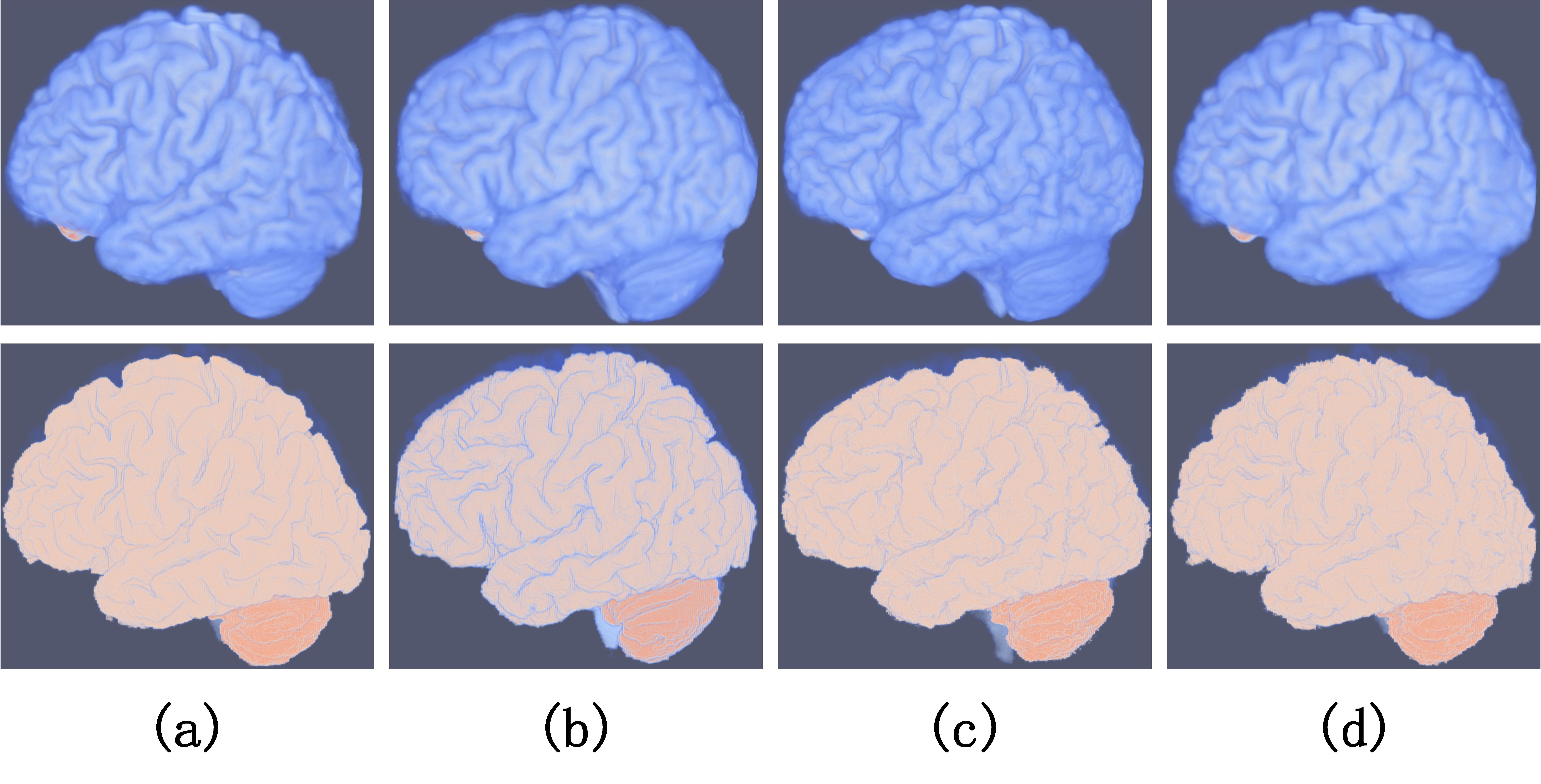}
  \vspace{-2pt}
  \caption{3D visualization for one pair of images processed by our DCCNN-LSTM-Reg. From top to bottom: Original image, Segmentation label; from left to right: (a) moving image $X$, (b) fixed image $Y$, (c) results of registration from $X$ to $Y$ ($\bm{X\rightarrow Y}$), (d) results of registration from $Y$ to $X$ ($\bm{X\leftarrow Y}$).}
  \label{fig:3D_image}
\end{figure}

Fig. \ref{fig:3D_image} shows 3D visualization of our DCCNN-LSTM-Reg registration for a pair of images. The forward and reverse registered results are very similar to the original images, and both have remarkable detail preservation.

\begin{figure*}[t!]
\centering
\includegraphics[width=\linewidth]{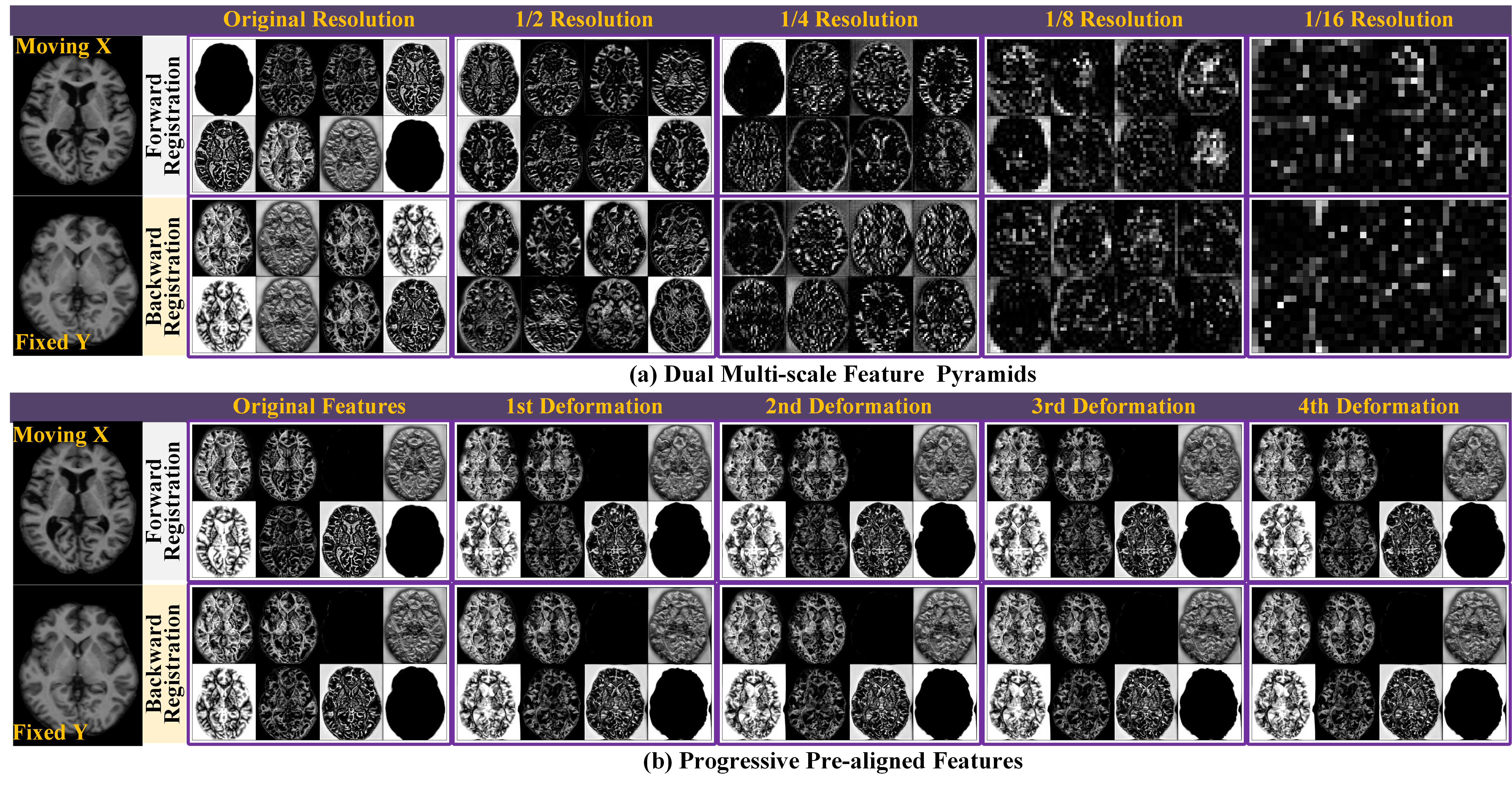}
\caption{(a) The feature maps in dual multi-scale feature pyramids, eight 2D slice feature maps are randomly selected from five scales in the two feature pyramids. (b) The feature maps during progressive registration at the last scale, the initial features in DCCNN-LSTM-Reg are gradually deformed during the registration process. }
\label{fig:features}
\end{figure*}

\begin{figure*}[t!]
\centering
\includegraphics[width=\linewidth]{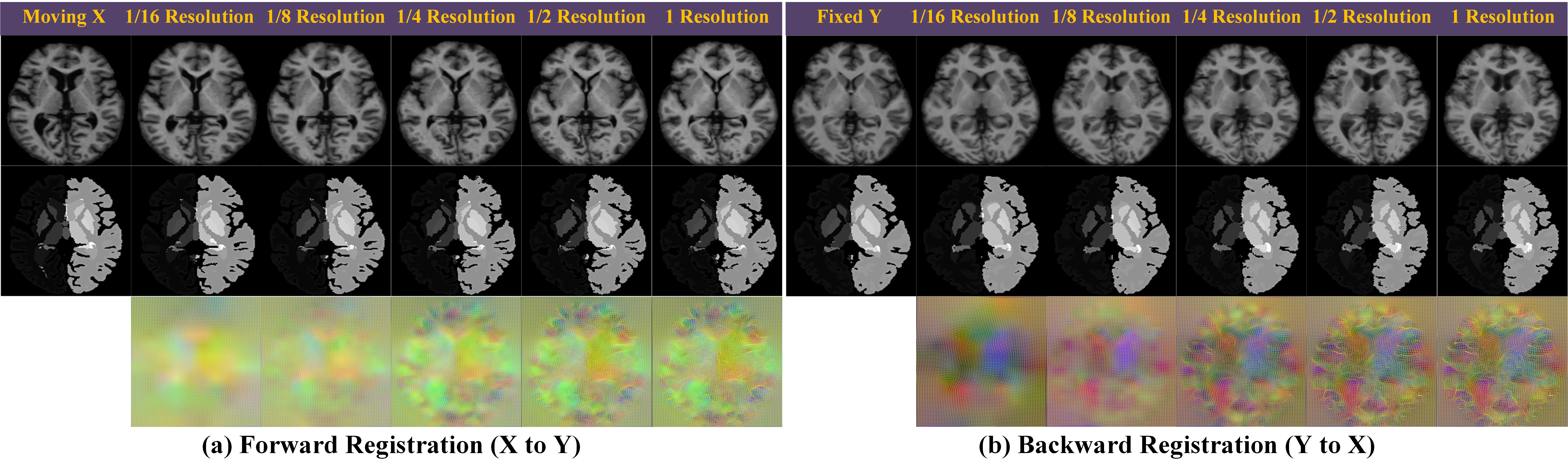}
\caption{Visualizations during multi-scale progressive registration in DCCNN-LSTM-Reg. From top to bottom: original image, segmentation labels, registration grid. From left to right in (a) and (b): original image, intermediate registered results at 1/16, 1/8, 1/4, 1/2, full resolution, respectively.}
\label{fig:scaleregpath}
\end{figure*}

\begin{figure*}[t!]
\centering
\includegraphics[width=\linewidth]{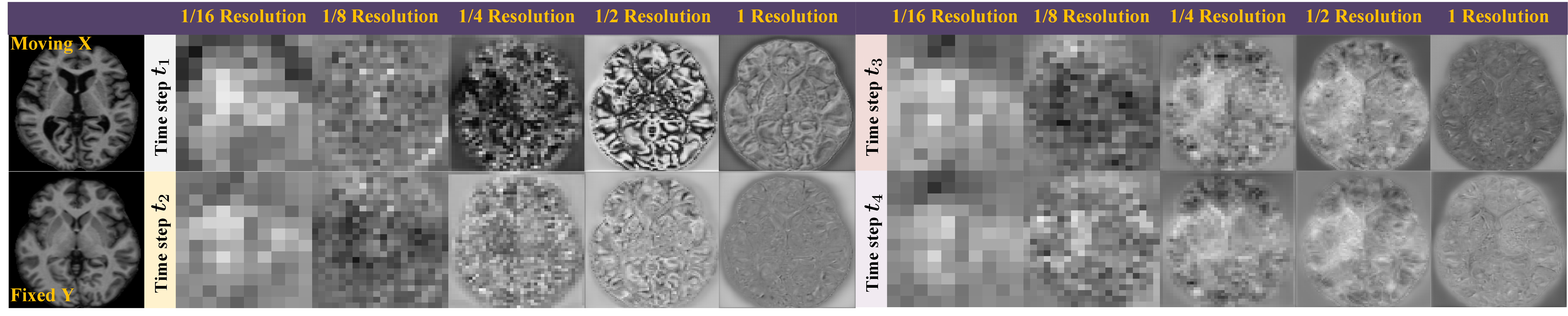}
\caption{Visualizations of memory features during multiscle progressive registration at multiple SR-module scales and different CNN-LSTM time steps, each image shows a gray level image obtained by averaging the values of the three channels of memory features $c$.
}
\label{fig:memory}
\end{figure*}

\subsection{Ablation Study}
We conducted extensive ablation experiments and analysis to assess the efficiency of each technical component of DCCNN-LSTM-Reg. The term S-DCCNN-LSTM-Reg is used to describe a simplified version, characterized by the removal of both the incremental control constraint loss for training and the progressive feature-deforming operation of the SR-Module in the original DCCNN-LSTM-Reg framework.

\subsubsection{Ablation Study On Network Components}
To begin, we evaluated the advantages of integrating the CNN-LSTM module, the U-net subnetwork, cycle consistency loss, smooth loss, Jacobian loss, and a symmetric registration strategy. In the S-DCCNN-LSTM-Reg framework, the SR-module is fed with the original features $\{(\bm{F}_X^\ell,\bm{F}_Y^\ell)\}_{\ell=1}^L$ (without the progressive deformation operation on the extracted features) and determines the deformation field by integrating the incremental field. Table \ref{tab: DCCNN-LSTM-Reg-ablation} presents the results of the ablation study conducted on the OASIS data set. By substituting LSTM with GRU in the CNN-LSTM model (variant (a)), the DSC scores were 0.794 and 0.704 for CNN-LSTM and CNN-GRU, respectively, indicating a 11\% improvement with CNN-LSTM. We then compared U-net with ResNet for feature extraction (variant (b)), with U-net outperforming ResNet by 1.5\%. Furthermore, replacing U-net with the feature extraction module from TransMorph (variant (c)) led to a reduction in the DSC score to 0.787. In terms of loss analysis, ablation experiments that omitted cycle consistency, smooth and Jacobian losses (variants (d)-(f)) showed different declines in DSC scores, and the increase in $\% |J_{\bm{\phi}}| \le0$ was prominent, reaching 1.6990 when Jacobian loss was omitted. Lastly, the change from symmetric to typical registration (variant (g)) caused a drop in the DSC score to 0.791. In particular, the value of $\% |J_{\bm{\phi}}| \le0$ increased by a factor of ten in the absence of a reverse registration strategy.

We started our investigation of small deformation constraint (control increment loss) and progressive feature-deforming operation using variants of S-DCCNN-LSTM-Reg (variants (h)-(j)). Incorporating the pre-aligned deformed features $[\bm{F}^\ell_X \circ (\bm{\phi}),\bm{F}^\ell_Y]$ and $[\bm{F}^\ell_X,\bm{F}^\ell_Y\circ (\bm{\phi}^{-1})]$ into CNN-LSTM (variant (h)) led to an accuracy increase to 0.809, even though the percentage of $\% J_{\bm{\phi}} \le0$ more than doubled. Including the small deformation constraint (variant (j)) did not affect the DSC average score, which stayed at 0.809, while the percentage of $\% J_{\bm{\phi}} \le0$ reduced to match the levels seen with the simplified DCCNN-LSTM-Reg.

\subsubsection{Ablation Study On Model Complexity}
We investigated how the complexity of a model affects the registration performance. Table \ref{tab: DCCNN-LSTM-Reg-OASIS-N} presents the results of S-DCCNN-LSTM-Reg with different cascade levels on the OASIS dataset. We can see from Table \ref{tab: DCCNN-LSTM-Reg-OASIS-N} that for $N=1$, S-DCCNN-LSTM-Reg does not significantly outperform the performance of other deep learning models. However, as $N$ increases, the image similarity score of S-DCCNN-LSTM-Reg also improves. Importantly, increasing the network layers does not cause significantly degradation, suggesting that the CNN-LSTM framework is well suited for multi-cascade registration. Iterative stacking enhances the model performance without leading to overfitting. As $N$ increases, the number of parameters also increases, reflecting a direct relationship. To balance model performance and computational cost, we therefore fix $N=4$ as the baseline for all subsequent experiments.

\subsubsection{Internal Network Visualization}
Fig. \ref{fig:features}(a) presents an illustration of the progressive feature-deforming volumes within dual multiscale feature pyramids, with eight 2D slices extracted at each scale. The features obtained closely match the theoretical forecasts. To enhance interpretability, two separate networks are utilized to extract features from the two images, rather than employing a U-net to extract features while learning the deformation field. At the coarse scale, the extracted features represent the global structure of the original image, whereas at the fine scale, they capture finer local details. Fig. \ref{fig:features}(b) shows the step-by-step deformation of features throughout multi-moment registration at the full scale. Starting with the initial features, they are deformed four times through the cascaded CNN-LSTM registration module, with each registration stage aligning the features appropriately.

The intermediate steps of multi-scale registration are depicted in Fig. \ref{fig:scaleregpath}. Beginning with a 1/16 resolution, the coarser scale helps approximate the deformation direction, while the finer scale refines the deformation of finer details. As the registration progresses, the source image aligns more closely with the fixed image, where the low-resolution grid shows only a general deformation, and higher-resolution grids acquire finer details with minimal folding.

Fig. \ref{fig:memory} illustrates the visualization of the memory feature $c$ in four distinct time steps on each scale within DCCNN-LSTM-Reg. The memory feature is composed of three channels, and the results shown are the average of these channels. Fig. \ref{fig:memory} clearly demonstrate that the memory feature starts at the coarsest scale in the registration path, capturing the registration process incrementally and refining it to achieve the final prediction. The memory feature conveys details at various levels, suggesting that the CNN-LSTM structure contributes to the registration path, aligning with theoretical expectations. This
combination of LSTM for registration is not only feasible, but also interpretable.

\section{Conclusion}\label{section:5}
This paper presents the DCCNN-LSTM-Reg framework, designed to improve the symmetrically diffeomorphic registration of adaptable medical images. Our innovative approach surpasses existing methods in effectiveness. The principal innovation of our work lies in integrating the deep learning framework with mathematical mechanisms of diffeomorphic image registration, allowing us to represent diffeomorphic registration as continuous transformations across multiple scales and time-dependent sequences. We tackle this challenge by utilizing homotopy continuation alongside progressive deformation fields that meet small deformation constraint (control increment loss). Through comprehensive experiments on three typical medical image registration tasks, we validate the superior performance of our method with quantitative and qualitative evaluations.

\bibliographystyle{IEEEtran}
\bibliography{DCC_Reg_diffeo_refs}

% Generated by IEEEtran.bst, version: 1.14 (2015/08/26)
\begin{thebibliography}{10}
\providecommand{\url}[1]{#1}
\csname url@samestyle\endcsname
\providecommand{\newblock}{\relax}
\providecommand{\bibinfo}[2]{#2}
\providecommand{\BIBentrySTDinterwordspacing}{\spaceskip=0pt\relax}
\providecommand{\BIBentryALTinterwordstretchfactor}{4}
\providecommand{\BIBentryALTinterwordspacing}{\spaceskip=\fontdimen2\font plus
\BIBentryALTinterwordstretchfactor\fontdimen3\font minus
  \fontdimen4\font\relax}
\providecommand{\BIBforeignlanguage}[2]{{%
\expandafter\ifx\csname l@#1\endcsname\relax
\typeout{** WARNING: IEEEtran.bst: No hyphenation pattern has been}%
\typeout{** loaded for the language `#1'. Using the pattern for}%
\typeout{** the default language instead.}%
\else
\language=\csname l@#1\endcsname
\fi
#2}}
\providecommand{\BIBdecl}{\relax}
\BIBdecl

\bibitem{YFu1}
Y.~Fu, Y.~Lei, T.~Wang, W.~J. Curran, T.~Liu, and X.~Yang, ``Deep learning in
  medical image registration: a review,'' \emph{Physics in Medicine \&
  Biology}, vol.~65, no.~20, p. 20TR01, 2020.

\bibitem{Yang2012}
X.~Yang, P.~Ghafourian, P.~Sharma, K.~Salman, D.~Martin, and B.~Fei, ``Nonrigid
  registration and classification of the kidneys in 3d dynamic contrast
  enhanced (dce) mr images,'' in \emph{Medical Imaging 2012: Image Processing},
  vol. 8314.\hskip 1em plus 0.5em minus 0.4em\relax SPIE, 2012, pp. 105--112.

\bibitem{Velec2011}
M.~Velec, J.~L. Moseley, C.~L. Eccles, T.~Craig, M.~B. Sharpe, L.~A. Dawson,
  and K.~K. Brock, ``Effect of breathing motion on radiotherapy dose
  accumulation in the abdomen using deformable registration,''
  \emph{International Journal of Radiation Oncology* Biology* Physics},
  vol.~80, no.~1, pp. 265--272, 2011.

\bibitem{Fu2011}
Y.~Fu, C.-K. Chui, C.~L. Teo, and E.~Kobayashi, ``Motion tracking and strain
  map computation for quasi-static magnetic resonance elastography,'' in
  \emph{Medical Image Computing and Computer-Assisted Intervention--MICCAI
  2011: 14th International Conference, Toronto, Canada, September 18-22, 2011,
  Proceedings, Part I 14}.\hskip 1em plus 0.5em minus 0.4em\relax Springer,
  2011, pp. 428--435.

\bibitem{Dang2014}
H.~Dang, A.~Wang, M.~S. Sussman, J.~Siewerdsen, and J.~Stayman, ``dpirple: a
  joint estimation framework for deformable registration and
  penalized-likelihood ct image reconstruction using prior images,''
  \emph{Physics in Medicine \& Biology}, vol.~59, no.~17, p. 4799, 2014.

\bibitem{JPZhang2015}
J.~Zhang and K.~Chen, ``Variational image registration by a total
  fractional-order variation model,'' \emph{J. Comput. Phys.}, vol. 293, pp.
  442--461, Jul. 2015.

\bibitem{demons1998}
J.-P. Thirion, ``Image matching as a diffusion process: an analogy with
  maxwell's demons,'' \emph{Medical image analysis}, vol.~2, no.~3, pp.
  243--260, 1998.

\bibitem{B1999}
D.~Rueckert, L.~I. Sonoda, C.~Hayes, D.~L. Hill, M.~O. Leach, and D.~J. Hawkes,
  ``Nonrigid registration using free-form deformations: application to breast
  mr images,'' \emph{IEEE transactions on medical imaging}, vol.~18, no.~8, pp.
  712--721, 1999.

\bibitem{LDDMM2005}
M.~F. Beg, M.~I. Miller, A.~Trouv{\'e}, and L.~Younes, ``Computing large
  deformation metric mappings via geodesic flows of diffeomorphisms,''
  \emph{International journal of computer vision}, vol.~61, pp. 139--157, 2005.

\bibitem{diffdemons2009}
T.~Vercauteren, X.~Pennec, A.~Perchant, and N.~Ayache, ``Diffeomorphic demons:
  Efficient non-parametric image registration,'' \emph{NeuroImage}, vol.~45,
  no.~1, pp. S61--S72, 2009.

\bibitem{syn2008}
B.~B. Avants, C.~L. Epstein, M.~Grossman, and J.~C. Gee, ``Symmetric
  diffeomorphic image registration with cross-correlation: evaluating automated
  labeling of elderly and neurodegenerative brain,'' \emph{Medical image
  analysis}, vol.~12, no.~1, pp. 26--41, 2008.

\bibitem{ZHANG2022}
J.~Zhang and Y.~Li, ``Diffeomorphic image registration with an optimal control
  relaxation and its implementation,'' \emph{SIAM Journal on Imaging Sciences},
  vol.~14, no.~4, pp. 1890--1931, 2021.

\bibitem{KCLam2014}
K.~C. Lam and L.~M. Lui, ``Landmark- and intensity-based registration with
  large deformations via quasi-conformal maps,'' \emph{SIAM Journal on Imaging
  Sciences}, vol.~7, no.~4, pp. 2364--2392, Jan. 2014.

\bibitem{CChen2018}
C.~Chen and O.~Öktem, ``Indirect image registration with large diffeomorphic
  deformations,'' \emph{SIAM Journal on Imaging Sciences}, vol.~11, no.~1, pp.
  575--617, Jan. 2018.

\bibitem{Daoping2018}
D.~Zhang and K.~Chen, ``A novel diffeomorphic model for image registration and
  its algorithm,'' \emph{J. Math. Imaging Vis.}, vol.~60, no.~8, pp.
  1261--1283, Apr. 2018.

\bibitem{CChen2019}
C.~Chen, B.~Gris, and O.~Öktem, ``A new variational model for joint image
  reconstruction and motion estimation in spatiotemporal imaging,'' \emph{SIAM
  Journal on Imaging Sciences}, vol.~12, no.~4, pp. 1686--1719, Jan. 2019.

\bibitem{HHan2020a}
H.~Han and Z.~Wang, ``A diffeomorphic image registration model with
  fractional-order regularization and {Cauchy}--{Riemann} constraint,''
  \emph{SIAM Journal on Imaging Sciences}, vol.~13, no.~3, pp. 1240--1271,
  2020.

\bibitem{Chen_2021}
C.~Chen, ``Spatiotemporal imaging with diffeomorphic optimal transportation,''
  \emph{Inverse Probl.}, vol.~37, no.~11, p. 115004, Oct. 2021.

\bibitem{HHan2021a}
H.~Han, Z.~Wang, and Y.~Zhang, ``Multiscale approach for two-dimensional
  diffeomorphic image registration,'' \emph{Multiscale Model. Simul.}, vol.~19,
  no.~4, pp. 1538--1572, 2021.

\bibitem{DZhang2022}
D.~Zhang, G.~P.~T. Choi, J.~Zhang, and L.~M. Lui, ``A unifying framework for
  n-dimensional quasi-conformal mappings,'' \emph{SIAM Journal on Imaging
  Sciences}, vol.~15, no.~2, pp. 960--988, Jun. 2022.

\bibitem{Alom2018}
M.~Z. Alom, T.~M. Taha, C.~Yakopcic, S.~Westberg, P.~Sidike, M.~S. Nasrin,
  B.~C. Van~Esesn, A.~A.~S. Awwal, and V.~K. Asari, ``The history began from
  alexnet: A comprehensive survey on deep learning approaches,'' \emph{arXiv
  preprint arXiv:1803.01164}, 2018.

\bibitem{Haskins2020}
G.~Haskins, U.~Kruger, and P.~Yan, ``Deep learning in medical image
  registration: a survey,'' \emph{Machine Vision and Applications}, vol.~31,
  pp. 1--18, 2020.

\bibitem{PXue2024Str}
P.~Xue, J.~Zhang, L.~Ma, M.~Liu, Y.~Gu, J.~Huang, F.~Liu, Y.~Pan, X.~Cao, and
  D.~Shen, ``Structure-aware registration network for liver dce-ct images,''
  \emph{IEEE Journal of Biomedical and Health Informatics}, vol.~28, no.~4, pp.
  2163--2174, 2024.

\bibitem{VoxelMorph2019}
G.~Balakrishnan, A.~Zhao, M.~R. Sabuncu, J.~Guttag, and A.~V. Dalca,
  ``Voxelmorph: a learning framework for deformable medical image
  registration,'' \emph{IEEE transactions on medical imaging}, vol.~38, no.~8,
  pp. 1788--1800, 2019.

\bibitem{diffVoxelMorph2018}
A.~V. Dalca, G.~Balakrishnan, J.~Guttag, and M.~R. Sabuncu, ``Unsupervised
  learning for fast probabilistic diffeomorphic registration,'' in
  \emph{Medical Image Computing and Computer Assisted Intervention--MICCAI
  2018: 21st International Conference, Granada, Spain, September 16-20, 2018,
  Proceedings, Part I}.\hskip 1em plus 0.5em minus 0.4em\relax Springer, 2018,
  pp. 729--738.

\bibitem{vtn2019}
S.~Zhao, Y.~Dong, E.~I. Chang, Y.~Xu \emph{et~al.}, ``Recursive cascaded
  networks for unsupervised medical image registration,'' in \emph{Proceedings
  of the IEEE/CVF international conference on computer vision}, 2019, pp.
  10\,600--10\,610.

\bibitem{CycleMorph2020}
B.~Kim, D.~H. Kim, S.~H. Park, J.~Kim, J.-G. Lee, and J.~C. Ye, ``Cyclemorph:
  cycle consistent unsupervised deformable image registration,'' \emph{Medical
  image analysis}, vol.~71, p. 102036, 2021.

\bibitem{sym2020}
T.~C. Mok and A.~Chung, ``Fast symmetric diffeomorphic image registration with
  convolutional neural networks,'' in \emph{Proceedings of the IEEE/CVF
  conference on computer vision and pattern recognition}, 2020, pp. 4644--4653.

\bibitem{RLiu2022}
R.~Liu, Z.~Li, X.~Fan, C.~Zhao, H.~Huang, and Z.~Luo, ``Learning deformable
  image registration from optimization: Perspective, modules, bilevel training
  and beyond,'' \emph{IEEE Transactions on Pattern Analysis and Machine
  Intelligence}, vol.~44, no.~11, pp. 7688--7704, 2022.

\bibitem{prnet2022}
M.~Kang, X.~Hu, W.~Huang, M.~R. Scott, and M.~Reyes, ``Dual-stream pyramid
  registration network,'' \emph{Medical image analysis}, vol.~78, p. 102379,
  2022.

\bibitem{Wei2022}
D.~Wei, S.~Ahmad, Y.~Guo, L.~Chen, Y.~Huang, L.~Ma, Z.~Wu, G.~Li, L.~Wang,
  W.~Lin \emph{et~al.}, ``Recurrent tissue-aware network for deformable
  registration of infant brain mr images,'' \emph{IEEE transactions on medical
  imaging}, vol.~41, no.~5, pp. 1219--1229, 2021.

\bibitem{YLiu2023Geo}
Y.~Liu, W.~Wang, Y.~Li, H.~Lai, S.~Huang, and X.~Yang, ``Geometry-consistent
  adversarial registration model for unsupervised multi-modal medical image
  registration,'' \emph{IEEE Journal of Biomedical and Health Informatics},
  vol.~27, no.~7, pp. 3455--3466, 2023.

\bibitem{AHering2023}
A.~Hering, L.~Hansen, T.~C.~W. Mok, A.~C.~S. Chung, H.~Siebert, S.~Häger,
  A.~Lange, S.~Kuckertz, S.~Heldmann, W.~Shao, S.~Vesal, M.~Rusu, G.~Sonn,
  T.~Estienne, M.~Vakalopoulou, L.~Han, Y.~Huang, P.-T. Yap, M.~Brudfors,
  Y.~Balbastre, S.~Joutard, M.~Modat, G.~Lifshitz, D.~Raviv, J.~Lv, Q.~Li,
  V.~Jaouen, D.~Visvikis, C.~Fourcade, M.~Rubeaux, W.~Pan, Z.~Xu, B.~Jian,
  F.~De~Benetti, M.~Wodzinski, N.~Gunnarsson, J.~Sjölund, D.~Grzech, H.~Qiu,
  Z.~Li, A.~Thorley, J.~Duan, C.~Großbröhmer, A.~Hoopes, I.~Reinertsen,
  Y.~Xiao, B.~Landman, Y.~Huo, K.~Murphy, N.~Lessmann, B.~van Ginneken, A.~V.
  Dalca, and M.~P. Heinrich, ``Learn2reg: Comprehensive multi-task medical
  image registration challenge, dataset and evaluation in the era of deep
  learning,'' \emph{IEEE Transactions on Medical Imaging}, vol.~42, no.~3, pp.
  697--712, 2023.

\bibitem{CLiu2024Reg}
C.~Liu, K.~He, D.~Xu, H.~Shi, H.~Zhang, and K.~Zhao, ``Regfsc-net: Medical
  image registration via fourier transform with spatial reorganization and
  channel refinement network,'' \emph{IEEE Journal of Biomedical and Health
  Informatics}, vol.~28, no.~6, pp. 3489--3500, 2024.

\bibitem{Arsigny2006}
V.~Arsigny, O.~Commowick, X.~Pennec, and N.~Ayache, ``A log-euclidean framework
  for statistics on diffeomorphisms,'' in \emph{Medical Image Computing and
  Computer-Assisted Intervention--MICCAI 2006: 9th International Conference,
  Copenhagen, Denmark, October 1-6, 2006. Proceedings, Part I 9}.\hskip 1em
  plus 0.5em minus 0.4em\relax Springer, 2006, pp. 924--931.

\bibitem{TransMorph2022}
J.~Chen, E.~C. Frey, Y.~He, W.~P. Segars, Y.~Li, and Y.~Du, ``Transmorph:
  Transformer for unsupervised medical image registration,'' \emph{Medical
  image analysis}, vol.~82, p. 102615, 2022.

\bibitem{QGChen2024}
Q.~Chen, Z.~Li, and L.~M. Lui, ``A deep learning framework for diffeomorphic
  mapping problems via quasi-conformal geometry applied to imaging,''
  \emph{SIAM Journal on Imaging Sciences}, vol.~17, no.~1, pp. 501--539, 2024.

\bibitem{HZHANG2024}
H.~Zhang and L.~M. Lui, ``A learning-based framework for topology-preserving
  segmentation using quasiconformal mappings,'' \emph{Neurocomputing}, vol.
  600, p. 128124, 2024.

\bibitem{Ashburner2007}
J.~Ashburner, ``A fast diffeomorphic image registration algorithm,''
  \emph{Neuroimage}, vol.~38, no.~1, pp. 95--113, 2007.

\bibitem{Vercauteren2009}
T.~Vercauteren, X.~Pennec, A.~Perchant, and N.~Ayache, ``Diffeomorphic demons:
  Efficient non-parametric image registration,'' \emph{NeuroImage}, vol.~45,
  no.~1, pp. S61--S72, 2009.

\bibitem{ConvLSTM2015}
X.~Shi, Z.~Chen, H.~Wang, D.-Y. Yeung, W.-K. Wong, and W.-c. Woo,
  ``Convolutional lstm network: A machine learning approach for precipitation
  nowcasting,'' \emph{Advances in neural information processing systems},
  vol.~28, 2015.

\bibitem{ncc2005}
B.~A. Ardekani, S.~Guckemus, A.~Bachman, M.~J. Hoptman, M.~Wojtaszek, and
  J.~Nierenberg, ``Quantitative comparison of algorithms for inter-subject
  registration of 3d volumetric brain mri scans,'' \emph{Journal of
  neuroscience methods}, vol. 142, no.~1, pp. 67--76, 2005.

\bibitem{oasis2007}
D.~S. Marcus, T.~H. Wang, J.~Parker, J.~G. Csernansky, J.~C. Morris, and R.~L.
  Buckner, ``Open access series of imaging studies (oasis): cross-sectional mri
  data in young, middle aged, nondemented, and demented older adults,''
  \emph{Journal of cognitive neuroscience}, vol.~19, no.~9, pp. 1498--1507,
  2007.

\bibitem{Hoopes2022}
A.~Hoopes, M.~Hoffmann, D.~N. Greve, B.~Fischl, J.~Guttag, and A.~V. Dalca,
  ``Learning the effect of registration hyperparameters with hypermorph,''
  \emph{The journal of machine learning for biomedical imaging}, vol.~1, p.
  003, March 2022.

\bibitem{freesurfer2012}
B.~Fischl, ``Freesurfer,'' \emph{Neuroimage}, vol.~62, no.~2, pp. 774--781,
  2012.

\bibitem{mindboggle101}
A.~Klein and J.~Tourville, ``101 labeled brain images and a consistent human
  cortical labeling protocol,'' \emph{Frontiers in neuroscience}, vol.~6, p.
  171, 2012.

\end{thebibliography}
\end{document}